\documentclass[pdflatex,sn-mathphys-num]{sn-jnl}

\usepackage{graphicx}%
\usepackage{multirow}%
\usepackage{amsmath,amssymb,amsfonts}%
\usepackage{amsthm}%
\usepackage{mathrsfs}%
\usepackage[title]{appendix}%
\usepackage[table]{xcolor}
\usepackage{textcomp}%
\usepackage{manyfoot}%
\usepackage{booktabs}%
\usepackage{algorithm}%
\usepackage{algorithmicx}%
\usepackage{algpseudocode}%
\usepackage{listings}%
\usepackage{tikz}
\usetikzlibrary{shapes.geometric, arrows}
\usepackage{array}
\usepackage{longtable}
\usepackage{tabularx}
\usepackage{lscape}
\usepackage[utf8]{inputenc}
\usepackage{paralist}
\usepackage{enumitem}
\usepackage{mathtools}
\usepackage{subcaption}
\usepackage{textgreek}
\usepackage{geometry}
\usepackage{siunitx}
\usepackage{float}
\usepackage{caption}
\usepackage{url}
\usepackage{tcolorbox} 
\usepackage{threeparttable}

\definecolor{claude}{HTML}{E69F00}
\definecolor{qwen}{HTML}{56B4E9}
\definecolor{chatgptO3}{HTML}{009E73}
\definecolor{gemini}{HTML}{D55E00}
\definecolor{deepseek}{HTML}{CC79A7}
\definecolor{mistral}{HTML}{999999}
\definecolor{grok}{HTML}{0072B2}
\definecolor{chatgpt4o}{HTML}{8C8C00}
\definecolor{llama}{HTML}{E76F51}

\tikzstyle{startstop} = [rectangle, rounded corners, minimum width=3cm, minimum height=1cm,text centered, draw=black, fill=red!30]
\tikzstyle{process} = [rectangle, minimum width=3cm, minimum height=1cm, text centered, draw=black, fill=orange!30]
\tikzstyle{decision} = [diamond, minimum width=3cm, minimum height=1cm, text centered, draw=black, fill=green!30]
\tikzstyle{arrow} = [thick,->,>=stealth]

\begin{document}

\title[Behavioral Augmentation of UML Class Diagrams]{Behavioral Augmentation of UML Class Diagrams: \\An Empirical Study of Large Language Models for Method Generation}

\author*{\fnm{ROUABHIA} \sur{Djaber}}\email{djaber.rouabhia@univ-tebessa.dz}

\author{\fnm{HADJADJ} \sur{Ismail}}\email{ismail.hadjadj@univ-tebessa.dz}

\affil{\orgdiv{Computer Science Department}, \orgname{Chahid Cheikh Laarbi Tebessi University}, \city{Tebessa}, \postcode{12000}, \state{Algeria}}

\abstract{
Automating the enrichment of UML class diagrams with behavioral methods from natural language use cases is a significant challenge. This study evaluates nine large language models (LLMs) in augmenting a methodless UML diagram (21 classes, 17 relationships) using 21 structured waste-management use cases. A total of 90 diagrams (3,373 methods) were assessed across six metrics: method quantity, signature richness (visibility, names, parameters, return types), annotation completeness (linking to use cases/actions), structural fidelity, syntactic correctness (PlantUML compilation), and naming convergence (across models). All LLMs produced valid PlantUML diagrams adhering to UML conventions. Some models excelled in method coverage and annotation accuracy, while others showed richer parameterization but weaker traceability. These results demonstrate that LLMs can generate well-structured methods with consistent naming, advancing automated behavioral modeling. However, inconsistencies in annotations and signatures highlight the need for improved prompt engineering and model selection. The rapid generation of these methods supports Agile practices by enabling faster design iterations. Despite their capabilities, human oversight is essential to ensure accuracy, appropriateness, and semantic alignment. This positions LLMs as collaborative partners in software design. All experimental artifacts (\texttt{.puml}, \texttt{.png}, \texttt{.csv}) are publicly available for reproducibility.
}


\maketitle
\newpage
\section{Introduction}  
\label{sec:intro}

Modern software engineering pipelines increasingly automate various development phases, such as testing, building, and deployment. However, the enrichment of UML class diagrams with behavioral details—such as method definitions, parameter specifications, return types, and traceability annotations—remains largely manual and error-prone. This manual effort often results in incomplete or inconsistent operational specifications, which can compromise design completeness and propagate defects into downstream artifacts such as API contracts, test scaffolds, and documentation. Consequently, software quality, maintainability, and traceability are negatively impacted, highlighting the critical need for more automated, reliable approaches to UML behavioral modeling.

Behavioral method modeling plays a pivotal role in accurately capturing system functionality and ensuring that software design aligns closely with intended use-case behaviors. Precise behavioral specifications improve communication among stakeholders, enhance model interpretability, and support downstream tasks such as code generation and testing. Despite its importance, automating behavioral enrichment in UML remains a challenging and underexplored problem.

Recent advances in large language models (LLMs) have demonstrated remarkable capacity to comprehend and generate natural language with high contextual awareness and semantic understanding. This linguistic proficiency positions LLMs as promising tools for bridging the gap between informal requirements and formal behavioral specifications. Specifically, their ability to interpret structured natural language use cases and infer corresponding method behaviors aligns well with the challenge of automatically enriching UML class diagrams. Leveraging these capabilities, LLMs could potentially automate the generation of detailed, well-annotated method definitions directly from textual requirements, addressing a critical bottleneck in early-stage software design. Additionally, LLMs have shown impressive capabilities in generating syntactically and semantically coherent code snippets, further suggesting their applicability to software design tasks.

While existing studies have demonstrated LLMs' potential in software design tasks, most of the work has focused on static aspects of UML class diagrams, such as class and attribute extraction. Behavioral method generation—especially automatic synthesis of methods from structured use cases—remains largely unaddressed. For instance, Saini et al.~\cite{Saini2022} extract classes and associations from user stories but omit operation synthesis, while Weyssow et al.~\cite{Weyssow2022} recommend metamodel concepts during interactive modeling but focus on attributes and relations. Zakaria et al.~\cite{Zakaria2025} show that transformer-based pipelines outperform heuristic NLP for static structure extraction, but behavioral details are neglected. Cámara et al.~\cite{Camara2023} applied ChatGPT to UML synthesis, revealing brittleness and weak traceability in method suggestions.

Building on these foundations, this work aims to fill the gap by systematically evaluating nine state-of-the-art LLMs—Claude~3.7, Gemini~2.5~Pro, Grok~3, Qwen~3, DeepSeek-R1, GPT-4o, ChatGPT-o3, Llama~4, and Mixtral~8×22B—on the task of automatic behavioral method generation for UML class diagrams. Unlike prior efforts that are limited to static structure extraction or qualitative assessments of single models, we provide a large-scale, multi-run, multi-model empirical benchmark that rigorously quantifies syntactic validity, structural fidelity, annotation completeness, and lexical consensus in the generated methods. Furthermore, all experimental artifacts, including prompts, enriched diagrams, and metric data, are publicly available.

To assess the models' effectiveness, we utilize a waste-management domain scenario, which captures common modeling challenges, such as diverse actors, use cases, and workflows. Although the findings are domain-specific, they offer valuable insights for other modeling scenarios, with future work planned to validate the generalizability of the results across diverse domains and UML modeling scenarios.

The core research question driving this study is:

\begin{quote}
\textbf{RQ1:} To what extent do LLMs generate syntactically valid UML methods from structured use-case descriptions that are well-formed, annotated, structurally consistent, and exhibit consensus across runs and models?
\end{quote}

Rather than defining a fixed benchmark, this study presents an empirical analysis focused on quantifiable syntactic, structural, and lexical aspects, acknowledging that semantic correctness and functional validation remain open challenges beyond the scope of this work.

The remainder of the paper is structured as follows: Section~\ref{sec:related} reviews related work; Section~\ref{sec:methodology} details the experimental setup; Sections~\ref{sec:results} and~\ref{sec:discussion} present and interpret the findings; and Section~\ref{sec:conclusion} concludes with future research directions.

\section{Related Work}  
\label{sec:related}

Research involving large language models (LLMs) in software design and code generation typically falls into three converging strands: (i) static UML artifact synthesis, (ii) conversational coding assistance, and (iii) benchmark-driven evaluation of generated code. Across these strands, the dynamic behavioral dimension of UML class diagrams—specifically, the automatic generation of methods including visibility markers, naming conventions, parameter richness, return types, and annotations—remains underexplored.

This gap is partly due to the inherent technical challenges in inferring method-level behavior from natural language, the lack of large, structured datasets linking use cases to detailed UML methods, and the historical focus on static structure extraction or code generation benchmarks that do not directly address UML behavioral semantics.

\subsection{Static UML Artifact Synthesis}

Early efforts in UML automation, such as Saini et al.~\cite{Saini2022}, focus on extracting classes, attributes, and associations from user stories, emphasizing traceability but stopping short of synthesizing operations. Weyssow et al.~\cite{Weyssow2022} use pretrained transformers to recommend metamodel concepts during interactive modeling sessions, concentrating on static elements like classes and relationships rather than behavioral enrichment. Similarly, Zakaria et al.~\cite{Zakaria2025} demonstrate that transformer-based pipelines outperform heuristic NLP methods for extracting static class diagrams but omit operation synthesis, which remains an unaddressed challenge. Cámara et al.~\cite{Camara2023} apply ChatGPT for UML synthesis, revealing brittle method suggestions and weak traceability, thus highlighting significant challenges in method generation and behavioral traceability.

In contrast to these studies, which primarily address static structure, this work expands the scope by systematically evaluating the automatic generation of methods—encompassing method signatures, annotations, and the consistency of method placement in UML class diagrams—using LLMs. Our focus on behavioral method generation from structured use cases represents a significant departure from previous efforts focused on static class extraction or isolated code generation tasks.

\subsection{Conversational Coding Assistance}

A second research direction treats LLMs as conversational programming assistants. Siam et al.~\cite{siam2024programmingAI} compare models like ChatGPT and GitHub Copilot for interactive coding tasks, revealing user preference for boilerplate code generation, although semantic errors persist in more complex code. While these systems improve productivity for ad-hoc code generation, they rarely evaluate the consistent generation of behavioral methods across models or handle the multi-method complexity inherent in class-level behavioral modeling. Unlike these interactive systems, which focus on isolated coding tasks, this study evaluates autonomous method synthesis for UML class diagrams, focusing on the full, automated enrichment of behavior, not just simple method generation.

\subsection{Benchmark-driven Evaluations of Code Generation}

Another strand of research evaluates the quality of code generated by LLMs through benchmark datasets such as HumanEval~\cite{chen2021codex} and MBPP~\cite{austin2021mbpp}, which measure single-function correctness via unit tests. More complex benchmarks, such as APPS~\cite{hendrycks2021apps} and AlphaCode~\cite{li2022alphacode}, extend to algorithmic challenges but still focus on isolated code generation. Du et al.~\cite{du2024classeval} introduce ClassEval, testing class implementations that require internal state and method interactions. Although ClassEval probes object-oriented reasoning, it is limited to Python and does not address UML modeling or cross-model method generation consensus.

In contrast to these benchmarks, which focus on isolated code generation and function-level accuracy, this study targets the automatic enrichment of UML class diagrams, evaluating LLMs on a multi-method, multi-model task that includes syntactic, structural, and behavioral consistency. This approach fills a significant gap by focusing on the underexplored problem of behavioral method generation for UML class diagrams, where the synthesis of methods—often omitted in static modeling benchmarks—plays a crucial role.

\subsection{Novelty and Contribution}

While much of the existing work has concentrated on static UML extraction, ad-hoc coding assistance, or function-level code generation, our study addresses the critical task of enriching UML diagrams with behavioral details automatically. By systematically evaluating nine state-of-the-art LLMs for their ability to generate well-formed methods from structured use cases, this work contributes a large-scale empirical benchmark that quantifies the LLM-generated methods along multiple dimensions: syntactic validity, structural integrity, annotation completeness, and lexical consensus.

This study represents a step forward in establishing an empirical foundation for trustworthy AI-assisted UML behavioral modeling, where previous works have either focused on partial UML enrichment or evaluated LLMs in isolation. The public release of all generated UML diagrams and associated metric datasets will serve as a valuable resource for reproducibility, comparison, and future research in automated software engineering.

As summarized in Table~\ref{tab:related-work-comparison}, prior studies have largely focused on static UML structure extraction or interactive modeling, while this study uniquely addresses the automatic generation of behavioral methods for UML class diagrams.

\begin{table}[ht]
\centering
\caption{Comparison of related works and the present study along key dimensions. Semantic correctness refers to formal or functional validation of method behavior.}
\label{tab:related-work-comparison}
\small
\begin{tabular}{p{3cm} p{3.5cm} p{3.5cm} p{3cm}}
\toprule
\textbf{Work} & \textbf{Focus} & \textbf{Behavioral Method Generation} & \textbf{Evaluation Scope} \\
\midrule
Saini et al.~\cite{Saini2022} & Static UML classes and associations & No & Traceability emphasis, no operation synthesis \\
Weyssow et al.~\cite{Weyssow2022} & Metamodel concept recommendations during interactive modeling & No & Qualitative recommendations on static elements \\
Zakaria et al.~\cite{Zakaria2025} & Transformer-based static class diagram extraction & No & Performance on static extraction only \\
Cámara et al.~\cite{Camara2023} & ChatGPT for UML synthesis & Limited & Method suggestions with weak traceability \\
Siam et al.~\cite{siam2024programmingAI} & Interactive coding assistants & No & Usability and correctness on ad-hoc code queries \\
HumanEval~\cite{chen2021codex}, MBPP~\cite{austin2021mbpp} & Single-function code generation benchmarks & No & Unit-test based correctness for functions \\
ClassEval~\cite{du2024classeval} & Object-oriented class implementation in Python & Yes (limited) & Functional correctness, language-specific, small scale \\
\midrule
\textbf{This Study} & Behavioral method enrichment for UML class diagrams & Yes & Syntactic, structural, lexical, and annotation evaluation; semantic correctness out of scope \\
\bottomrule
\end{tabular}
\end{table}

\section{Methodology}  
\label{sec:methodology}

This study evaluates the ability of nine large language models (LLMs) to enrich a UML class diagram—initially devoid of methods—based on a structured set of use-case (UC) descriptions and a standardized prompt. The work follows a \emph{generate–measure–compare} paradigm~\cite{wohlin2012}, targeting a single research question (RQ1): 

\bigskip
\emph{To what extent do LLMs generate methods from structured use-case descriptions written in natural language that are syntactically valid UML elements, produce well-formed and detailed signatures, include the required inline annotations, preserve the structural integrity of the original class diagram, and exhibit consensus across runs and models?}

\begin{figure}[ht]
 \centering
 \includegraphics[width=\linewidth]{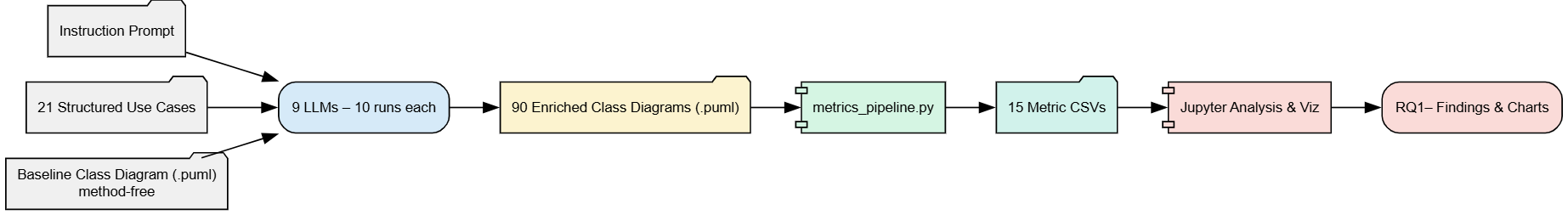}
 \caption{End-to-end pipeline from input prompts to enriched diagrams and metrics. The process involves standardized prompting, LLM method generation, UML diagram augmentation, parsing, and multi-dimensional evaluation.}
 \label{fig:pipeline}
\end{figure}

\bigskip
Figure~\ref{fig:pipeline} illustrates the overall pipeline from input prompts through to enriched diagrams and metric computation.

\subsection{Artifacts}

Several key artifacts form the basis of this evaluation:

\begin{itemize}
  \item \textbf{LLM Prompt:} A standardized, three-part prompt designed to guide method-level content generation aligned with domain behavior. It comprises (1) primary instructions specifying enrichment and annotation requirements, (2) the baseline UML class diagram in PlantUML format (a widely used textual UML diagram notation), and (3) the complete set of structured use cases in natural language. 

  \item \textbf{Baseline Diagram:} A methodless UML class model representing a waste recycling system, consisting of 21 classes and 17 relationships, serving as the foundation for method enrichment.

  \item \textbf{Use-Case Corpus:} A collection of 21 structured use cases, each including a title, ID, preconditions, postconditions, triggers, main scenario steps, and possible extensions. These are expressed in consistent natural language format covering core domain behaviors.

  \item \textbf{Augmented PlantUML Diagrams:} Ninety enriched UML diagrams generated across ten independent runs per model. These diagrams embed methods with visibility markers, method names, typed parameter lists, return types, and inline use-case and action annotations. Files follow the naming pattern \texttt{ModelName\_RunX.puml} (e.g., \texttt{Claude3\_run7.puml}).

  \item \textbf{Graphical Renderings:} PNG images compiled from each augmented diagram, named identically to the \texttt{.puml} files (e.g., \texttt{Claude3\_run7.png}).

  \item \textbf{Parsed JSON Files:} Structured representations extracted programmatically from the augmented PlantUML diagrams, supporting all metric calculations. Files follow the \texttt{ModelName\_RunX.json} format.

  \item \textbf{Metric CSV Files:} Tabular summaries of quantitative metrics derived from JSON analyses, covering syntactic, structural, lexical, annotation, and consensus dimensions.

\end{itemize}

All artifacts—including baseline diagrams, use cases, augmented \texttt{.puml} files, their graphical \texttt{.png} renderings, parsed \texttt{.json} files, and metric \texttt{.csv} outputs—are publicly accessible in the project repository (\url{https://github.com/Djaber1979/-Behavioral-Augmentation-of-UML-Class-Diagrams}).

\subsection{Prompting Method}

To accommodate model-specific constraints such as maximum prompt length and input limits, a consistent three-prompt strategy was adopted. This approach balanced feasibility and uniformity across diverse LLM architectures.

\begin{enumerate}[leftmargin=*]
  \item \textbf{Primary Instruction Prompt:} Provided detailed instructions directing the LLM to enrich a UML class diagram with method definitions derived from structured use cases. The LLM was instructed to produce a complete, syntactically valid PlantUML diagram with inline annotations linking each method to its source use case and specific scenario action. The full prompt text is available in Appendix~\ref{appendix:prompt}.

  \item \textbf{Diagram Prompt:} Supplied the baseline UML class diagram in PlantUML format, excluding methods. This model is publicly available in the repository at: \url{<https://github.com/Djaber1979/-Behavioral-Augmentation-of-UML-Class-Diagrams/blob/main/Prompts/Methodless.txt}.

  \item \textbf{Use-Case Prompt:} Supplied all 21 structured use cases in a unified input block. Each use case followed a consistent format with title, ID, preconditions, postconditions, trigger, main scenario steps, and extensions. The full set is available at: \url{<https://github.com/Djaber1979/-Behavioral-Augmentation-of-UML-Class-Diagrams/blob/main/Prompts/UCs.txt>}.
\end{enumerate}

This tripartite design was chosen after discarding two alternatives: prompting each use case individually exceeded input count limits in some models, while merging all inputs into a single prompt exceeded token limits in others. The adopted structure ensured input consistency and cross-model compatibility.

Each model was restricted to a single-shot generation without clarifications, feedback, or retries. Outputs were saved verbatim for downstream analysis.

\paragraph{Prompt-driven constraints.}
The final prompt imposed structured constraints to ensure domain-relevant and consistent method generation:
\begin{itemize}
\item \textbf{Source of actions:} The LLM was instructed to extract methods exclusively from the \texttt{MainScenario} sections of use cases, explicitly omitting alternative flows and edge cases to prioritize primary domain behavior—aligned with standard use-case modeling practices~\cite{cockburn2000writing,larman2004applying,jacobson1992usecase}

\item \textbf{Atomicity:} The LLM was instructed that “1 method = 1 atomic business action,” requiring decomposition of compound steps (e.g., those using “and”) to uphold single-responsibility principles (SRP).

\item \textbf{Domain-only scope:} The LLM was restricted to core business logic. Infrastructure and UI-related behaviors (e.g., \texttt{save()}, \texttt{display()}) were explicitly excluded.

\item \textbf{Naming:} Clear, domain-specific verbs (e.g., \texttt{activateaccount}) were encouraged; vague or generic labels (e.g., \texttt{setStatus}) were discouraged to ensure semantic clarity.

\item \textbf{Class assignment:} Methods were to be added only to existing UML classes, and assigned to the most semantically appropriate one. New class creation was disallowed.

\item \textbf{Traceability annotations:} Each method included inline comments linking it to its originating use case ID (\texttt{//UCXX}) and specific action step (\texttt{//action:}) for traceability.

\item \textbf{Explicit state transitions:} State changes were to be expressed through named methods (e.g., \texttt{confirmOrder()}), avoiding generic setters and embedding state logic directly in behavior, supporting single-responsibility and expressive modeling.
\end{itemize}

\subsection{Models}

Nine publicly accessible large language models (LLMs) were evaluated in this study, spanning diverse architectures, parameter scales, and training paradigms. These models include both dense and mixture-of-experts (MoE) architectures, each with varying parameter counts, context lengths, and training focuses. Notably, two distinct variants of OpenAI’s ChatGPT, \textbf{GPT-4o} (May 2024) and \textbf{ChatGPT-o3} (April 2025), were treated separately due to significant architectural and behavioral differences. Additionally, the evaluation covers models from various providers, including Claude, Qwen, Gemini, DeepSeek, Mistral, and Llama, each with unique strengths in reasoning, coding, and multimodal tasks. For clarity, the detailed specifications of each evaluated model, including version, release date, context length, architecture, and parameter counts (where available), are provided in Appendix~\ref{appendix:llm-summary} (Table~\ref{tab:llm-summary}).

\subsection{Metrics for RQ1}

To comprehensively address RQ1, seven complementary metric categories were designed, targeting key facets of UML method enrichment:

\begin{itemize}[leftmargin=*]
  \item \textbf{Method Quantity (MQ):} Total method counts per diagram, run variability, and class distribution patterns.
  \item \textbf{Signature Richness (SR):} Detailed method signature elements including visibility, parameters, return types, and lexical diversity.
  \item \textbf{Annotation Completeness (AC):} Presence of use-case and action-level annotations within methods.
  \item \textbf{Structural Fidelity (SF):} Preservation of original UML structure, measured via element-wise Jaccard similarity.
  \item \textbf{Syntactic Correctness (SC):} Rate of diagrams compiling without PlantUML syntax errors.
  \item \textbf{Top-Method Consensus (TMC):} Lexical alignment of frequent method names across models.
  \item \textbf{Core-Method Consensus and Structural Placement Consistency (CMC + SPC):} Agreement on generation and placement of core methods across runs and models.
\end{itemize}

\paragraph{Top-Method Consensus (TMC)}

Top-Method Consensus (TMC) assesses lexical convergence across LLMs without relying on a fixed reference set. All method names generated across models and runs are pooled and normalized, (e.g., lowercased, punctuation-stripped). The \(k\) most frequent names are selected, with \(k = \left\lceil \frac{\textit{total methods}}{\textit{total diagrams}} \right\rceil\).

This set captures the most salient, spontaneously reproduced behaviors, independent of structural context or class assignment.

Inter-model similarity is quantified by computing pairwise \textbf{Jaccard similarities} between raw (unnormalized) method name sets of each model.

Together with core-method consensus, TMC provides a lens into naming behavior: the former based on reproduction of a common core set, the latter on unsupervised lexical convergence.

\paragraph{Core-Method Consensus (CMC)}

To complement TMC, Core-Method Consensus identifies a fixed core set of frequently generated methods. Method names across all augmented diagrams are pooled and normalized, then the \textbf{top-\(k\)} most frequent method names are extracted, where \(k\) equals the overall mean number of methods per diagram (\(k \approx 38\)), reflecting typical model output size.

For each LLM, we compute:

\begin{itemize}[leftmargin=*]
  \item \textbf{Core-method coverage:} Proportion of the top-\(k\) core methods present in the model's outputs;
  \item \textbf{Method-level agreement:} Number or percentage of models that independently generate each core method.
\end{itemize}

This scheme captures the degree to which models reproduce a shared high-frequency behavioral vocabulary.

\paragraph{Structural Placement Consistency (SPC)}

To extend the analysis of core methods with a structural dimension, the stability of structural placement measures whether LLM assign core methods to their \emph{dominant class}, defined as the class most frequently associated with each method in all diagrams.

The \emph{Class Match Rate} for an LLM is computed as:

\[
\text{ClassMatch}_{\text{LLM}} = \frac{|\{m \in \text{Core}_k \mid m \text{ assigned to dominant class}\}|}{|\{m \in \text{Core}_k \mid m \in \text{LLM output}\}|}
\]

This metric reflects inter-model agreement not only on which methods to generate but also on where structurally they belong, indicating deeper behavioral and architectural consensus.

These metrics collectively operationalize RQ1 by quantitatively evaluating the breadth, detail, traceability, structural integrity, syntactic validity, and naming consistency of methods generated by LLMs. Semantic correctness and functional validation are recognized as very important but remain outside the current scope, slated for a future investigation.

\subsection{Statistical Procedures}

To robustly analyze metric differences across LLMs, nonparametric statistical methods were employed:

\begin{itemize}[leftmargin=*]
  \item \textbf{Kruskal–Wallis Tests} for detecting global between-model differences without assuming normality.
  \item \textbf{Dunn’s Post-Hoc Tests} with Holm correction to identify pairwise differences controlling for multiple comparisons.
  \item \textbf{Cliff’s \(\Delta\)} to estimate effect sizes and quantify the magnitude of observed differences.
\end{itemize}

Effect size thresholds follow Romano et al.~\cite{romano2006cliffs} for Cliff’s \(\Delta\) (small = 0.147, medium = 0.33, large = 0.474) and Cohen~\cite{cohen1988statistical} for Cramér’s \(V\) (small = 0.1, medium = 0.3, large = 0.5), providing practical context beyond mere significance.

Lexical metrics such as Levenshtein diversity and Jaccard similarity were analyzed at the run level, while coverage and placement consistency metrics aggregated at model or method granularity.

Stability of results was ensured via 1,000-sample bootstrap resampling for confidence estimation. All experiments were repeated across three independent seeds (17, 42, 123) to mitigate sampling bias, consistent with recent LLM benchmarking best practices~\cite{chiang2024chatbotarena} and following multi-seed evaluation protocols common in recent NLP studies (e.g., \cite{Dodge2020FineTuning,li2022evaluating}). Sensitivity analyses confirmed stability of main findings across seeds. All \(p\)-values were corrected using the Holm method within each metric family. Outliers were visualized but retained following evaluation practices~\cite{arcuri2014}.

Full metric definitions and detailed calculation procedures are available in Appendix~\ref{sec:appendix-metrics} and Appendix~\ref{sec:appendix-stats}.

\subsection{Sample Size and Justification}

Each of the nine LLMs was evaluated across ten independent runs, yielding 90 enriched class diagrams. This sampling strategy balances computational feasibility with sufficient coverage of stochastic output variability inherent in generative models.

The total corpus of 3,373 generated methods enables statistically meaningful analysis of method generation patterns and consensus.

Lexical convergence metrics (TMC, CMC, SPC) were computed using frequency-based method sets derived from the entire output corpus, allowing unified behavioral vocabulary comparisons without requiring predefined benchmarks.

\subsection{Scope and Empirical Nature}

This study adopts an empirical evaluation framework focused on syntactic correctness, structural fidelity, lexical patterns, and annotation completeness in LLM-generated UML behavioral enrichments. These aspects were selected for their objectivity and reproducibility across models and runs.

Due to the inherent ambiguity in mapping natural language use cases to behavioral method specifications—and the lack of a universally accepted ground truth—semantic and functional correctness are not evaluated in this study. Instead, they are intentionally deferred to future work.

Planned directions for semantic evaluation include expert-based assessments, formal validation techniques, embedding-based similarity analysis (e.g., cosine similarity), and automated behavioral testing. These approaches will enable deeper insights into the alignment between generated methods and intended domain logic.

This study thus establishes a quantitative baseline, supporting future semantic investigations by first grounding analysis in measurable, structural characteristics.

\section{Results}  
\label{sec:results}

This section presents results for \textbf{RQ1}, evaluating how effectively nine publicly available LLMs enrich structural UML class diagrams by generating methods derived from 21 structured use cases.

Evaluation covers nine core dimensions: six quantitative metrics—Method Quantity (MQ), Signature Richness (SR), Annotation Completeness (AC), Structural Fidelity (SF), Syntactic Correctness (SC)—and three lexical agreement metrics—Top-Method Consensus (TMC), Core-Method Consensus (CMC), and Structural Placement Consistency (SPC). Detailed metric definitions and formulas are provided in Appendix~\ref{sec:appendix-metrics} and Appendix~\ref{sec:appendix-stats}.

Across all runs, LLMs processed inputs—including methodless UML class diagrams, structured use cases, and enrichment instructions—to generate fully updated PlantUML code in times ranging from a few seconds to under five minutes. This rapid end-to-end processing significantly surpasses typical human capabilities for manual enrichment, underscoring the practical feasibility and efficiency of LLM-assisted UML behavioral augmentation in real-world software engineering workflows.

Statistics summarize \num{90} generated diagrams comprising \num{3,373} methods. Table~\ref{tab:results-overview} provides an overview of evaluation dimensions, the highest-scoring models per metric based on mean scores, and the corresponding statistical tests. Effect sizes were considered alongside significance to assess practical impact.

\begin{table}[ht]
\centering
\caption{Overview of evaluation dimensions: highest-scoring model(s) by metric, statistical test used, test statistic, and significance level.}
\label{tab:results-overview}
\begin{tabular}{p{1.5cm}p{3.5cm}p{2.5cm}p{2cm}p{1.2cm}}
\toprule
\textbf{Dimension} 
  & \textbf{Top-Scoring Model(s)} 
  & \textbf{Test} 
  & \textbf{Statistic} 
  & \(\mathbf{p}\)\textbf{-value} \\

\midrule
MQ 
  & Claude~3.7 
  & Kruskal–Wallis 
  & \(H(8)=51.85\) 
  & \(<.0001\) \\

SR 
  & Mixtral~8×22B 
  & Kruskal–Wallis 
  & \(H(8)=383.21\) 
  & \(<.0001\) \\

AC 
  & Claude~3.7, ChatGPT-o3 
  & Chi-squared 
  & \(\chi^2(16)=3001.07\) 
  & \(<.0001\) \\

SF 
  & Claude~3.7, ChatGPT-o3, Mistral 
  & Mixed ANOVA (model × run) 
  & \(F_{\text{model}}(8,81)=7.78\) 
  & \(<.0001\) \\

SC 
  & No significant differences 
  & Chi-squared 
  & \(\chi^2(8)=6.43\) 
  & \(0.5993\) \\

TMC 
  & Qwen~3 
  & Kruskal–Wallis 
  & \(H(8)=26.42\) 
  & \(0.0009\) \\

CMC 
  & Claude~3.7 
  & Kruskal–Wallis 
  & \(H(8)=26.42\) 
  & \(0.0009\) \\

SPC 
  & — 
  & Wilcoxon signed-rank 
  & \(W=123\) 
  & \(0.0112\) \\
\bottomrule
\end{tabular}
\end{table}
\noindent\textit{Note:} The models listed achieved the highest mean scores within each evaluation dimension; these should not be interpreted as universally “best” models, since project priorities vary. Reported \(p\)-values are corrected for multiple comparisons using Bonferroni or Holm methods. Model comparisons consider both statistical significance and effect size.

\begin{figure}[ht]
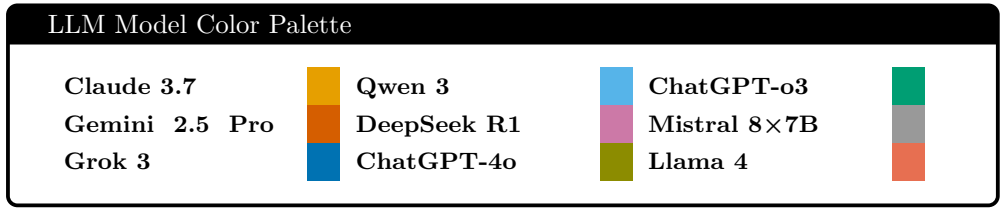

\centering
\begin{tcolorbox}[title=LLM Model Color Palette, colback=white, colframe=black]
\small
\renewcommand{\arraystretch}{1.3}
\begin{tabular}{>{\raggedright}p{3cm} c 
                >{\raggedright}p{3cm} c 
                >{\raggedright}p{3cm} c}
\textbf{Claude~3.7}      & \cellcolor{claude} &
\textbf{Qwen~3}          & \cellcolor{qwen} &
\textbf{ChatGPT-o3}      & \cellcolor{chatgptO3} \\

\textbf{Gemini~~2.5~~Pro}& \cellcolor{gemini} &
\textbf{DeepSeek~R1}     & \cellcolor{deepseek} &
\textbf{Mistral~8×7B}    & \cellcolor{mistral} \\

\textbf{Grok~3}          & \cellcolor{grok} &
\textbf{ChatGPT-4o}      & \cellcolor{chatgpt4o} &
\textbf{Llama~4}         & \cellcolor{llama} \\
\end{tabular}
\end{tcolorbox}
\caption{Color assignments used for LLM model visualizations.}
\label{tab:llm-palette}
\end{figure}

For consistent interpretation across visualizations, each LLM is assigned a fixed color throughout this article (see Figure~\ref{tab:llm-palette}).

Findings are interpreted considering both statistical significance and practical implications for software engineering workflows, such as maintainability, interpretability, and traceability in UML behavioral augmentation.

\subsection{Method Quantity (MQ)}

All nine LLMs generated UML methods across all runs, producing a total of \textbf{3,373 methods} over 90 diagrams. Table~\ref{tab:mq-summary} summarizes the total methods generated, per-run minimums and maximums, and syntax error occurrences for each model.

\begin{table}[ht]
\centering
\caption{Method generation volume and syntax correctness per model.}

\label{tab:mq-summary}
\small
\begin{tabular}{lrrrr}
\toprule
\textbf{Model} & \textbf{Total Methods} & \textbf{Min/Run} & \textbf{Max/Run} & \textbf{Syntax Errors} \\
\midrule
Claude~3.7                 & 734 & 63  & 85  & 0 \\
Gemini~2.5~Pro             & 477 & 20  & 79  & 2 \\
ChatGPT-o3                 & 432 & 25  & 61  & 0 \\
Qwen~3                     & 373 & 28  & 58  & 0 \\
DeepSeek~R1                & 316 & 14  & 42  & 1 \\
Grok~3                     & 294 & 8   & 72  & 1 \\
ChatGPT-4o                 & 275 & 18  & 33  & 1 \\
Llama~4                    & 268 & 17  & 40  & 1 \\
Mistral~8×7B               & 204 & 18  & 23  & 0 \\
\midrule
\textbf{Total (All Models)} & \textbf{3,373} &     &     & \textbf{6} \\
\bottomrule
\end{tabular}
\end{table}

\begin{figure}[ht]
  \centering
  \includegraphics[width=0.75\linewidth]{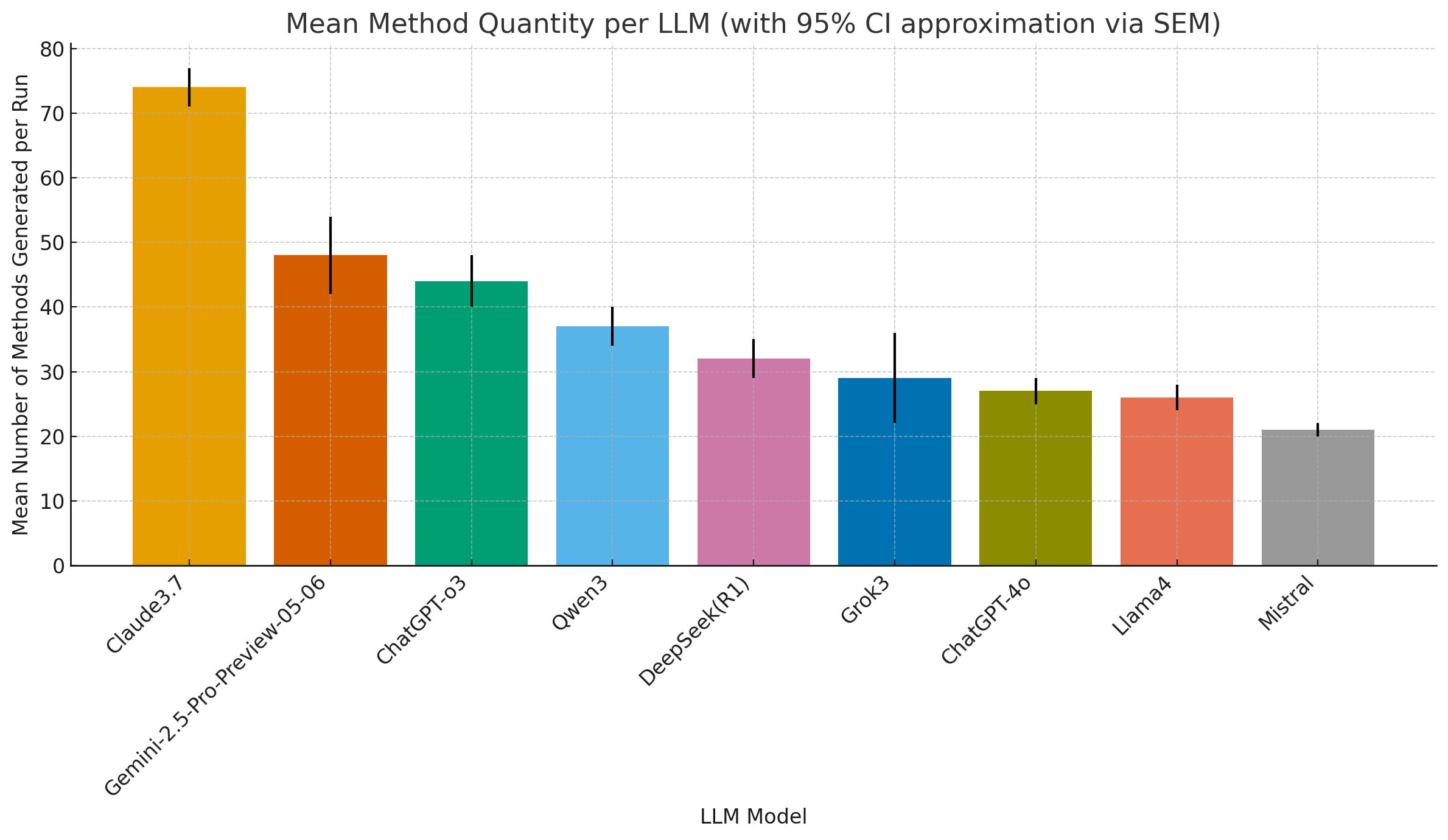}
  \caption{Mean Method Quantity per LLM with 95\% Confidence Intervals (SEM).}
  \label{fig:mq-bar}
\end{figure}

\begin{figure}[ht]
  \centering
  \includegraphics[width=0.75\linewidth]{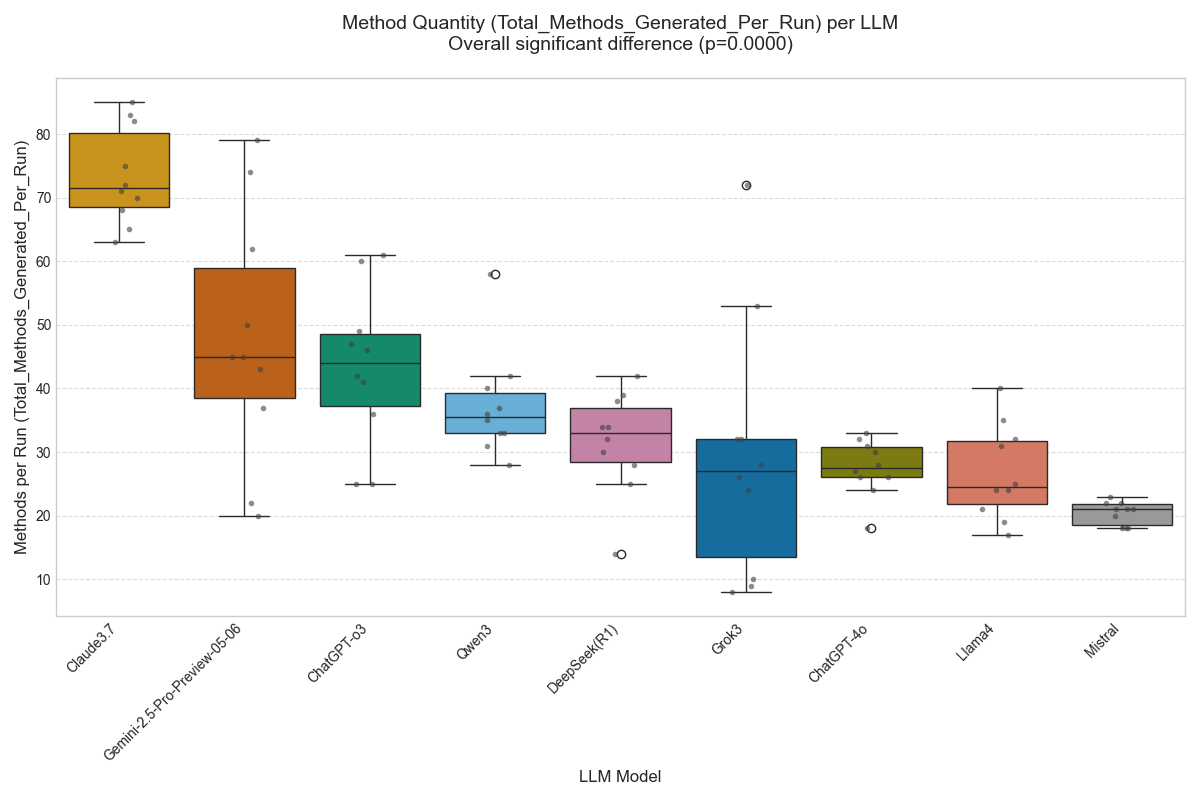}
  \caption{Per-Run Distributions of Total Methods Generated Across 10 Runs.}
  \label{fig:mq-boxplot}
\end{figure}

Figure~\ref{fig:mq-bar} displays the mean method quantity per LLM with 95\% confidence intervals (SEM)\footnote{SEM (Standard Error of the Mean)}, while Figure~\ref{fig:mq-boxplot} illustrates per-run distributions of total methods generated across the 10 runs. 

A Kruskal–Wallis test revealed statistically significant differences in method counts across models (\(H=51.85, p < .0001\)).

Post-hoc Dunn’s tests identified \textbf{Claude~3.7} as producing significantly more methods than \textbf{Grok~3}, \textbf{Llama~4}, and \textbf{Mistral}, with large effect sizes (e.g., rank-biserial correlation \(r_{\mathrm{rb}} = -1.00\) compared to Mistral). Additionally, \textbf{Gemini~2.5~Pro} significantly outperformed \textbf{Mistral}.

\begin{figure}[ht]
  \centering
  \includegraphics[width=\linewidth]{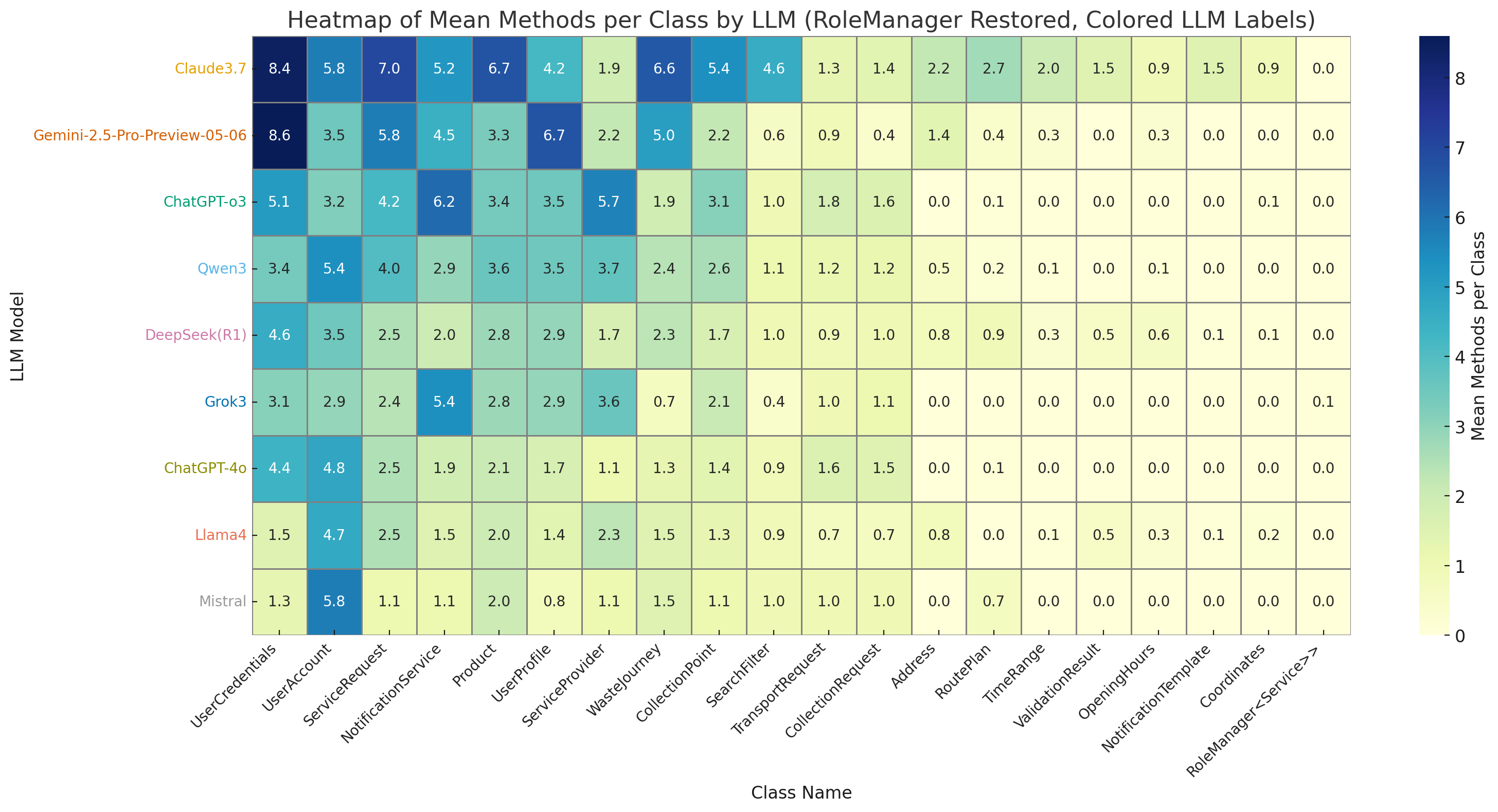}
  \caption{Heatmap of mean methods per UML class grouped by LLM. Higher values indicate more frequent method placement in the corresponding class.}
  \label{fig:mq-heatmap}
\end{figure}

\begin{figure}[ht]
  \centering
  \includegraphics[width=0.85\linewidth]{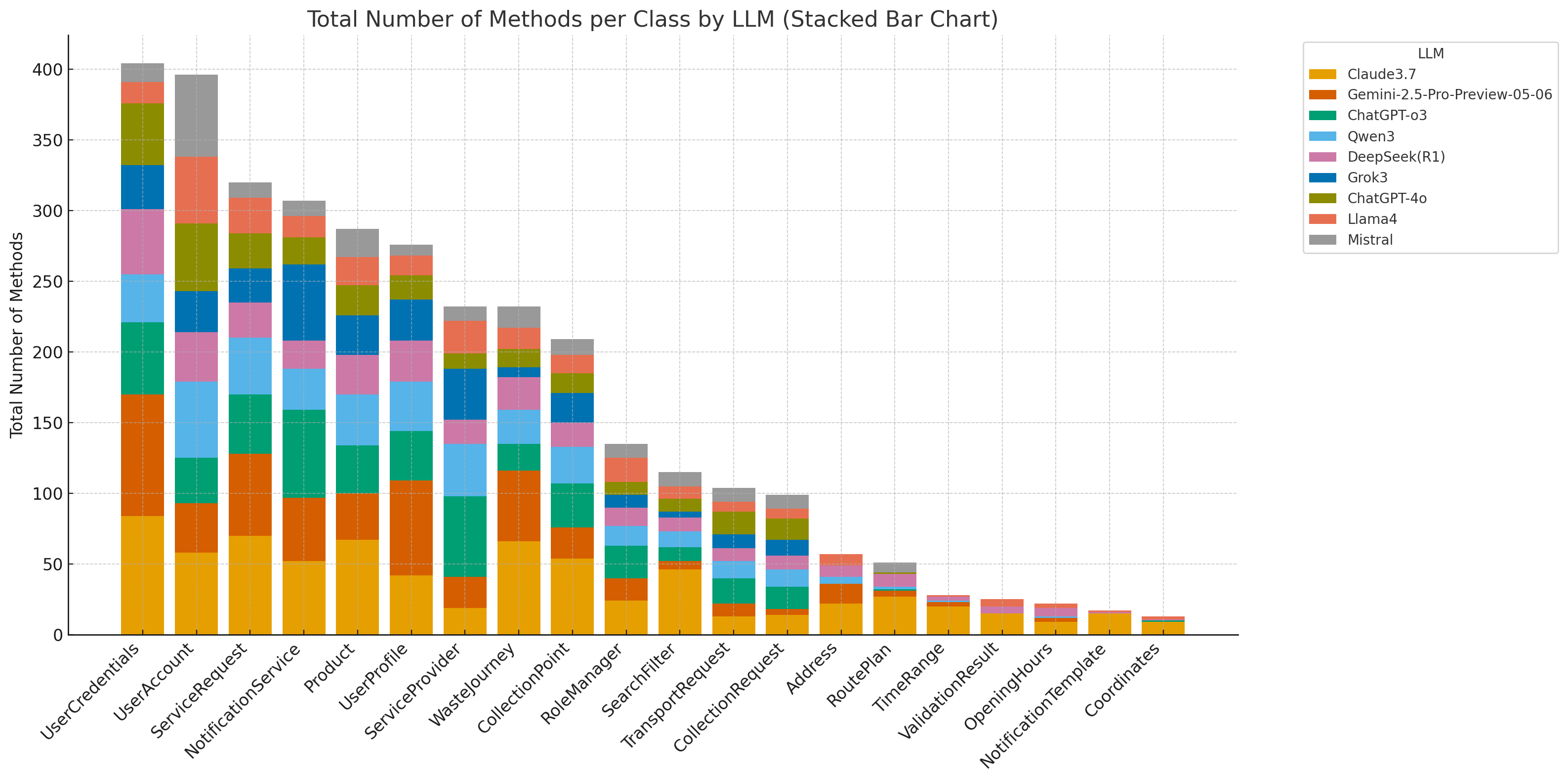}
  \caption{Method distribution per class aggregated across all runs and LLMs. Each bar represents a UML class, with colored segments indicating the contributions of each LLM using the consistent color palette (Table~\ref{tab:llm-palette}). Classes are ordered by descending total method count.}
  \label{fig:mq-method-distribution}
\end{figure}

Heatmaps and bar charts (Figures~\ref{fig:mq-heatmap} and \ref{fig:mq-method-distribution}) illustrate class-level method allocation patterns and aggregate method distributions across models. For example, Claude~3.7 distributes methods more evenly across core classes, indicating its suitability for comprehensive behavioral enrichment.

\paragraph{Interpretation and Practical Implications}  
Claude~3.7’s substantially higher method generation suggests it is well-suited for projects requiring comprehensive behavioral coverage, enabling detailed capture of domain-specific functionality. Conversely, Mistral’s more conservative output favors simplicity and maintainability, advantageous in contexts prioritizing clarity and ease of understanding. This variability across models highlights the importance of balancing extensive method coverage with practical downstream usability in software engineering workflows. Moreover, the consensus in class-level method placement supports LLMs’ reliability in identifying domain-relevant classes for behavioral augmentation. The low number of syntax errors—only six total (6.67\%)—confirms the high syntactic quality of generated code, facilitating smooth integration into UML tooling pipelines.

\subsection{Signature Richness (SR)}

Signature richness was evaluated across five dimensions: visibility markers, naming conventions, parameter richness, return types, and lexical coherence. Table~\ref{tab:sr-metrics} summarizes key metrics per LLM.

\begin{table}[ht]
  \centering
  \caption{Signature richness per LLM.}
  \label{tab:sr-metrics}
  \small
  \begin{tabular}{lccccc}
    \toprule
    \textbf{Model} & \textbf{IQR} & \textbf{Ret.} & \textbf{LexDiv} & \textbf{+} & \textbf{-- / \# / None} \\
    \midrule
    Claude~3.7     & 1.00 & 54.50 & 0.800 & 100.00 & 0.00 / 0.00 / 0.00 \\
    Gemini~2.5~Pro & 1.00 & 33.54 & 0.791 & 98.32  & 0.42 / 1.26 / 0.00 \\
    ChatGPT‑o3     & 1.25 & 39.58 & 0.801 & 100.00 & 0.00 / 0.00 / 0.00 \\
    Qwen~3         & 1.00 & 26.54 & 0.780 & 99.73  & 0.27 / 0.00 / 0.00 \\
    DeepSeek~R1    & 1.00 & 45.25 & 0.777 & 99.37  & 0.00 / 0.00 / 0.63 \\
    Grok~3         & 1.00 & 35.71 & 0.750 & 99.66  & 0.34 / 0.00 / 0.00 \\
    ChatGPT‑4o     & 2.00 & 35.27 & 0.802 & 100.00 & 0.00 / 0.00 / 0.00 \\
    Llama~4        & 0.00 & 12.31 & 0.807 & 100.00 & 0.00 / 0.00 / 0.00 \\
    Mistral        & 3.00 & 77.45 & 0.799 & 100.00 & 0.00 / 0.00 / 0.00 \\
    \bottomrule
  \end{tabular}
\end{table}

Here, \textbf{IQR} quantifies the variability in parameters per method, while \textbf{Ret} is the proportion of non-void return types (\%), and \textbf{LexDiv}—the normalized Levenshtein distance among unique method names—captures naming variation and expressiveness.

All models predominantly use the public visibility marker (`+`), with a statistically significant but small association between model and visibility usage (\(\chi^2 = 64.09, p < .0001\), Cramér's \(V = 0.080\)). Naming conventions are uniformly camelCase (100\%), reflecting adherence to the main prompt instructions and industry-standard practice.

\begin{figure}[ht]
  \centering
  \includegraphics[width=0.7\linewidth]{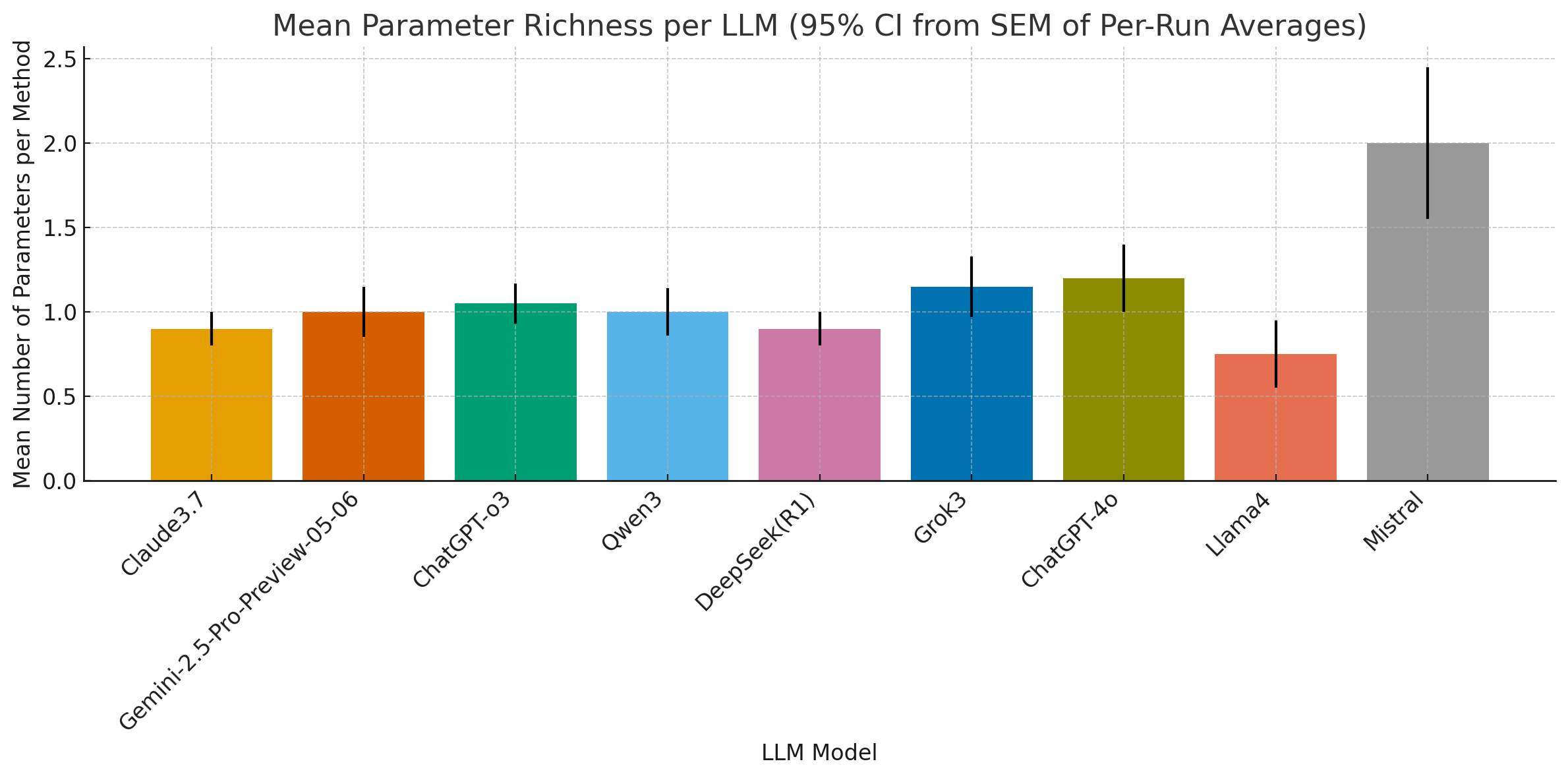}
  \caption{Mean parameter richness per method by LLM. Error bars show 95\% confidence intervals (SEM).}
  \label{fig:sr-param-bar}
\end{figure}

\begin{figure}[ht]
  \centering
  \includegraphics[width=0.7\linewidth]{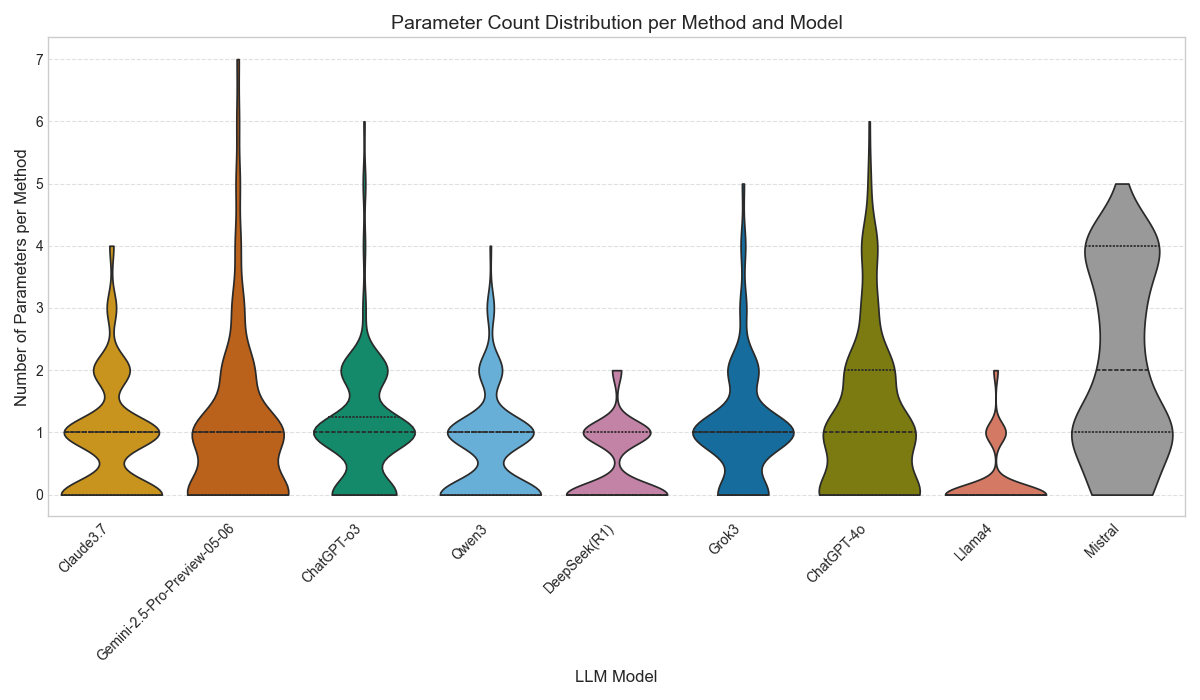}
  \caption{Distribution of parameter counts per method by model. Width indicates frequency; dashed lines mark quartiles.}
  \label{fig:sr-param-violin}
\end{figure}

Parameter richness varies significantly across models (\(H = 382.08, p < .0001\)), with Mistral generating notably richer parameter lists supported by large effect sizes (Cliff’s \(\delta = 1.0\) vs.\ Claude~3.7). Return-type annotations also differ (\(\chi^2 = 312.38, p < .0001\)), with Mistral and Claude~3.7 exhibiting higher proportions of non-void returns (Figures~\ref{fig:sr-param-bar}, \ref{fig:sr-param-violin}).

\begin{figure}[ht]
  \centering
  \includegraphics[width=0.9\linewidth]{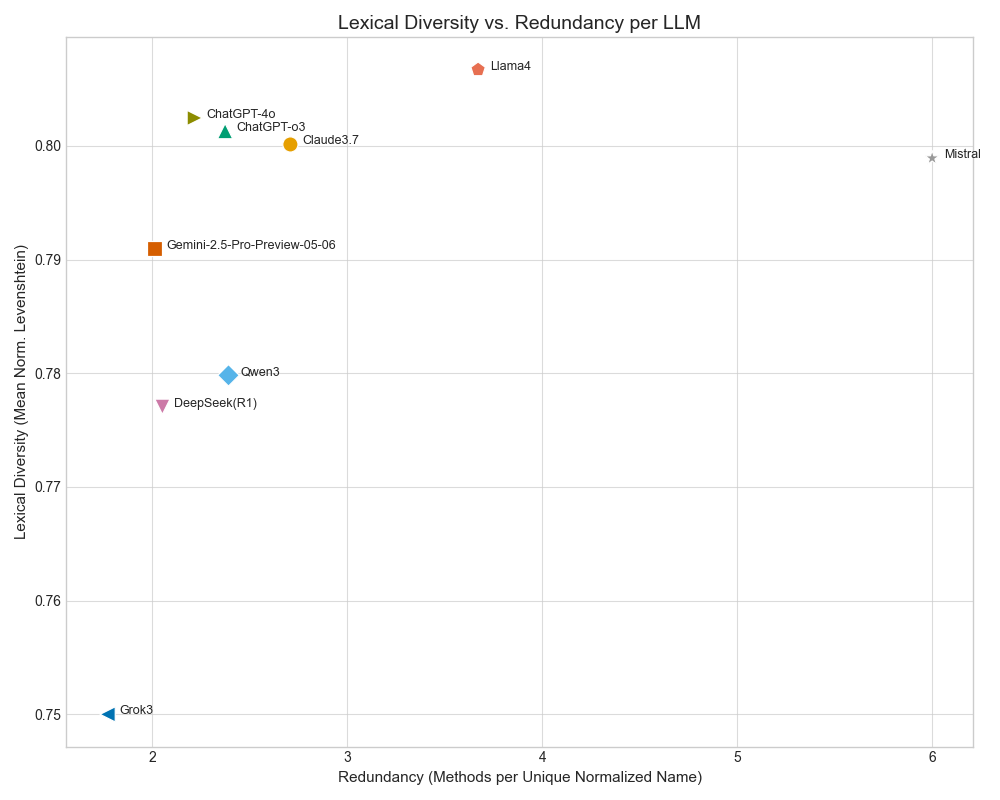}
  \caption{Lexical coherence versus redundancy across models. Y-axis: normalized Levenshtein distance; X-axis: method-name redundancy.}
  \label{fig:sr-lexdiv-scatter}
\end{figure}

Lexical diversity, measured by normalized Levenshtein distance, also varies significantly (\(H = 44.07, p < .0001\)). Claude~3.7 and ChatGPT variants exhibit higher lexical diversity, indicating more expressive naming, whereas Grok~3 displays more conservative naming (Figure~\ref{fig:sr-lexdiv-scatter}).

\paragraph{Interpretation and Practical Implications}  
Variation in signature richness reflects differing model design philosophies. Mistral’s richer parameter and return-type annotations enhance behavioral detail but may increase complexity. Models with greater lexical diversity tend to produce clearer, more maintainable method names. These findings highlight the importance of aligning LLM selection with project goals that balance detail and clarity.

\subsection{Annotation Completeness (AC)}

\begin{figure}[ht]
  \centering
  \includegraphics[width=0.75\linewidth]{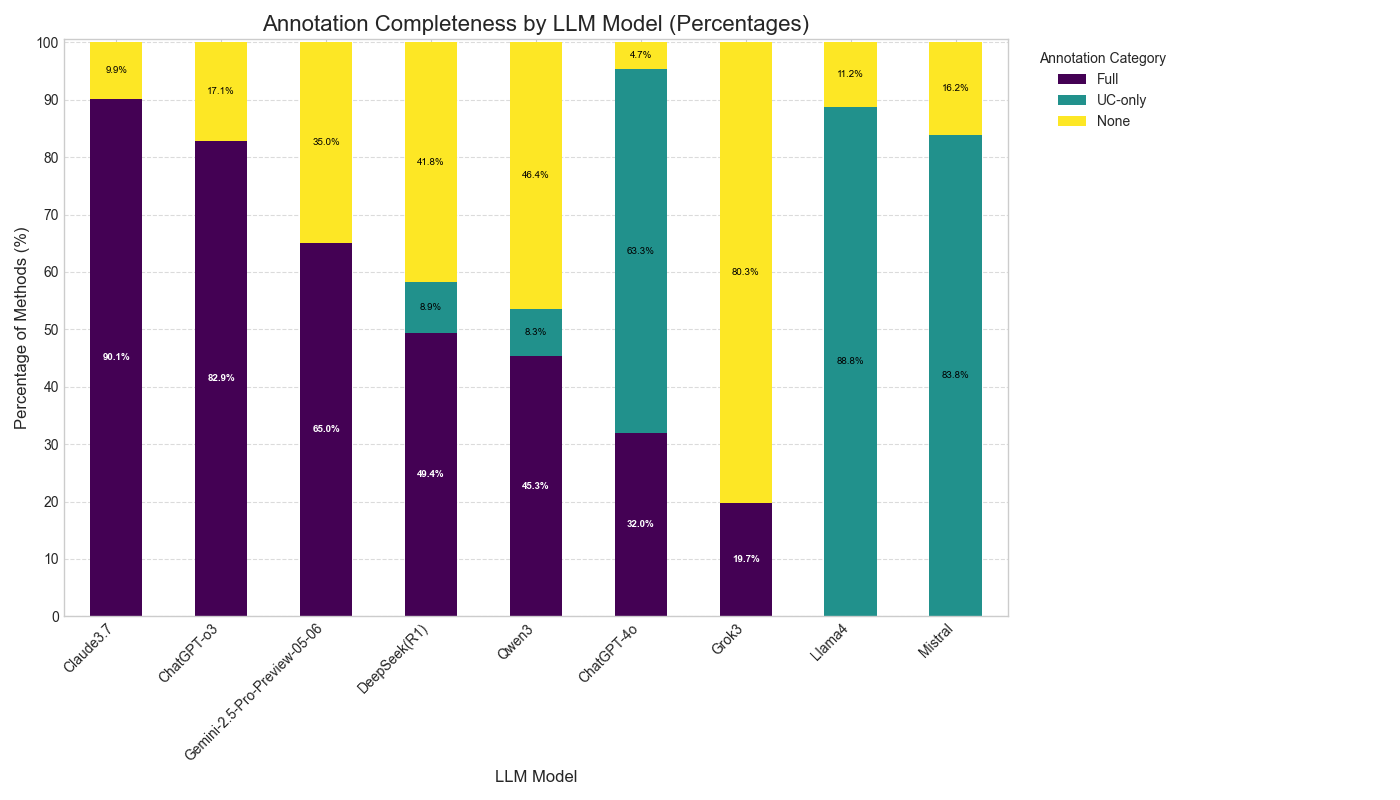}
  \caption{Annotation completeness by model.}
  \label{fig:ac-stacked-bar}
\end{figure}

Annotation completeness varies substantially across models. Figure~\ref{fig:ac-stacked-bar} illustrates the proportions of methods with full annotations (both \texttt{//UCXX} and \texttt{//action:}), use-case-only annotations, or no annotations.

\begin{figure}[ht]
  \centering
  \includegraphics[width=0.75\linewidth]{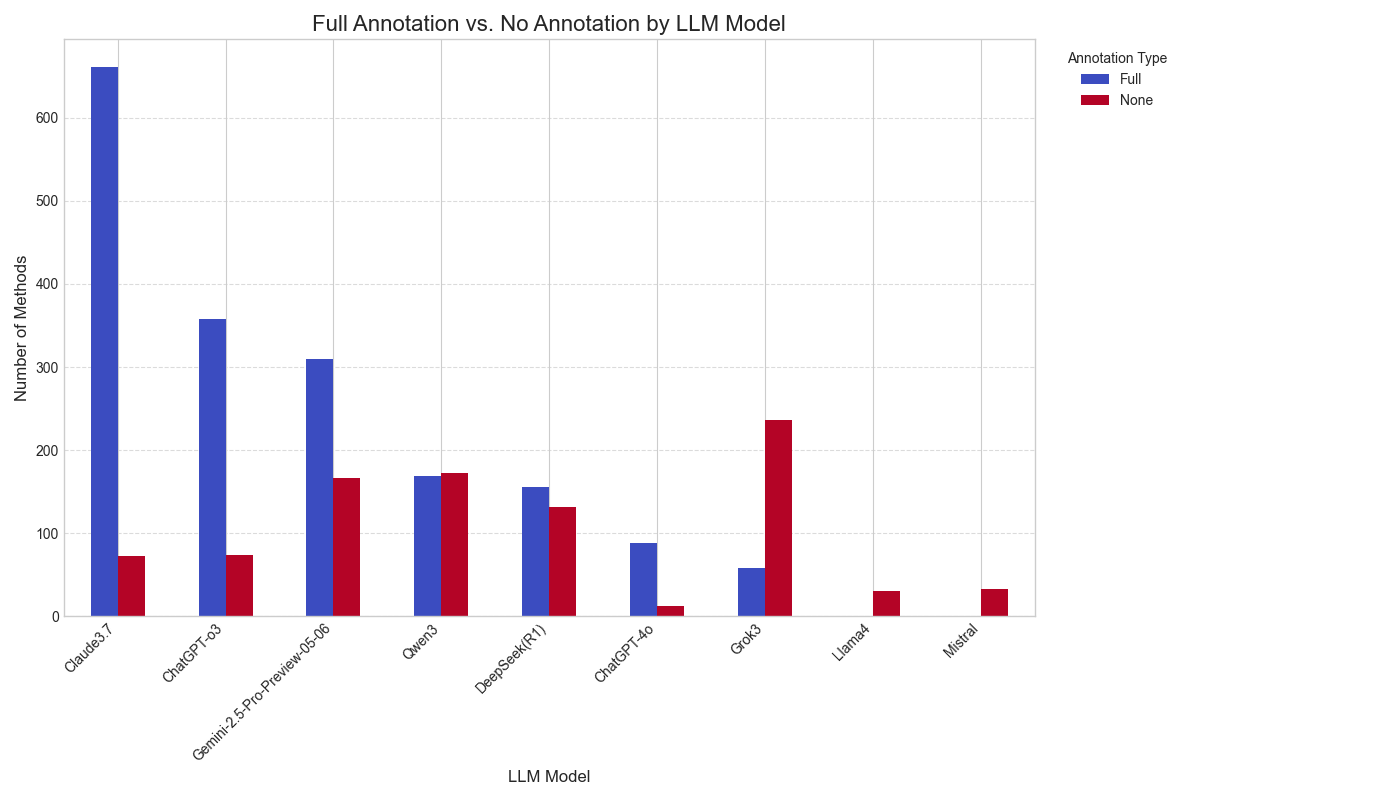}
  \caption{Counts of fully annotated versus unannotated methods by model.}
  \label{fig:ac-bar-full-none}
\end{figure}

Models such as \textbf{Claude~3.7} and \textbf{ChatGPT-o3} exhibit superior annotation adherence, enhancing traceability critical for maintainability and team collaboration. Conversely, \textbf{Mistral} and \textbf{Llama~4} frequently omit annotations, which may limit artifact usefulness. A strong association between model identity and annotation completeness was confirmed (\(\chi^2(16) = 3001.07, p < .0001\), Cramér’s \(V = 0.667\)), indicating that annotation quality is a consistent, model-specific characteristic (see Figure~\ref{fig:ac-bar-full-none}).

\paragraph{Interpretation and Practical Implications}  
Annotation completeness is essential for producing maintainable UML artifacts. High annotation fidelity facilitates traceability and supports collaborative software engineering workflows, suggesting it should be a key criterion when selecting LLMs for behavioral augmentation. Models with lower annotation rates may necessitate additional refinement or post-processing to ensure artifact utility.

\subsection{Structural Fidelity (SF)}

\begin{figure}[ht]
  \centering
  \includegraphics[width=\linewidth]{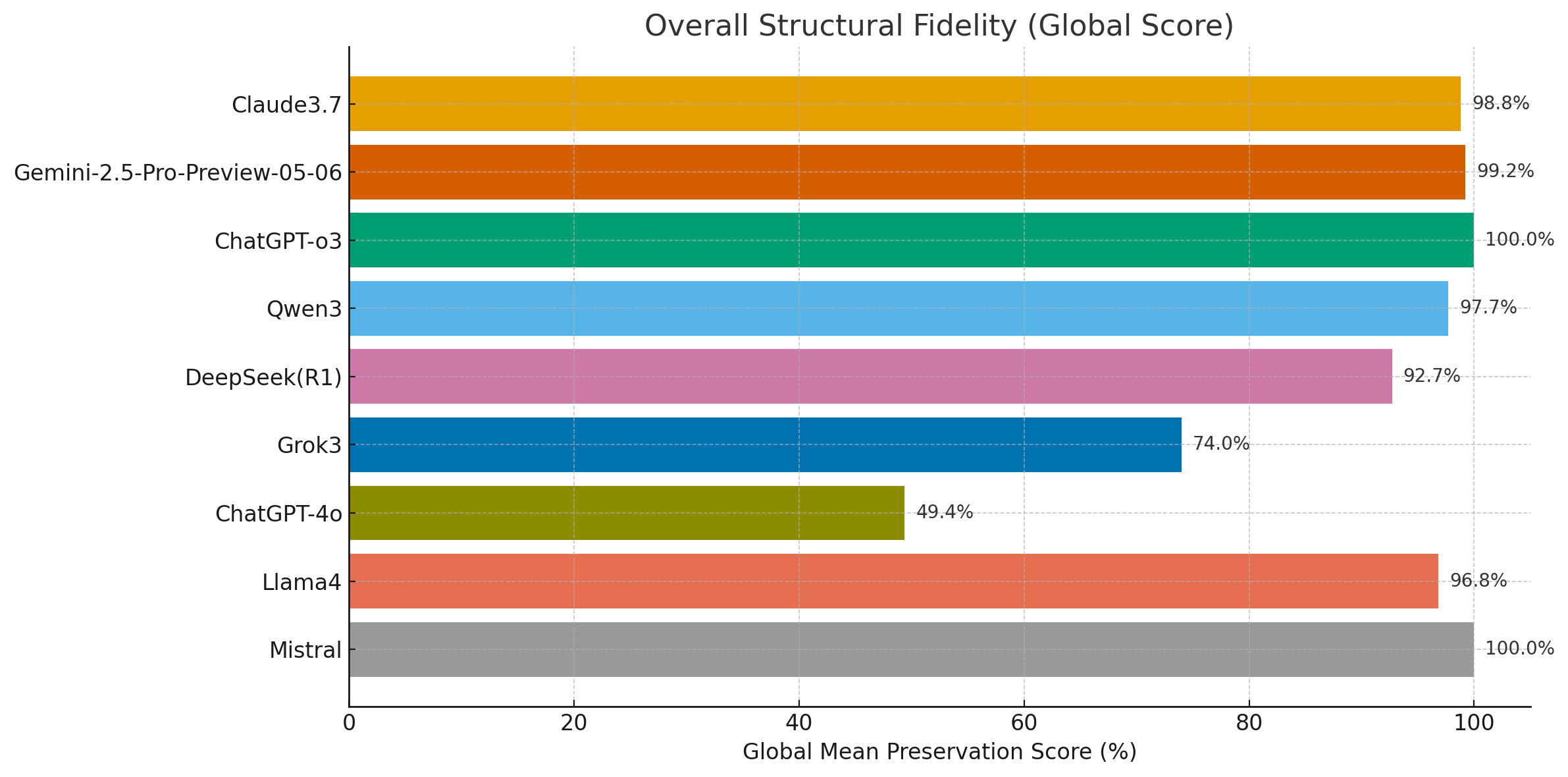}
  \caption{Average preservation across six UML elements per model.}
  \label{fig:sf-bar}
\end{figure}

\begin{figure}[ht]
  \centering
  \includegraphics[width=0.6\linewidth]{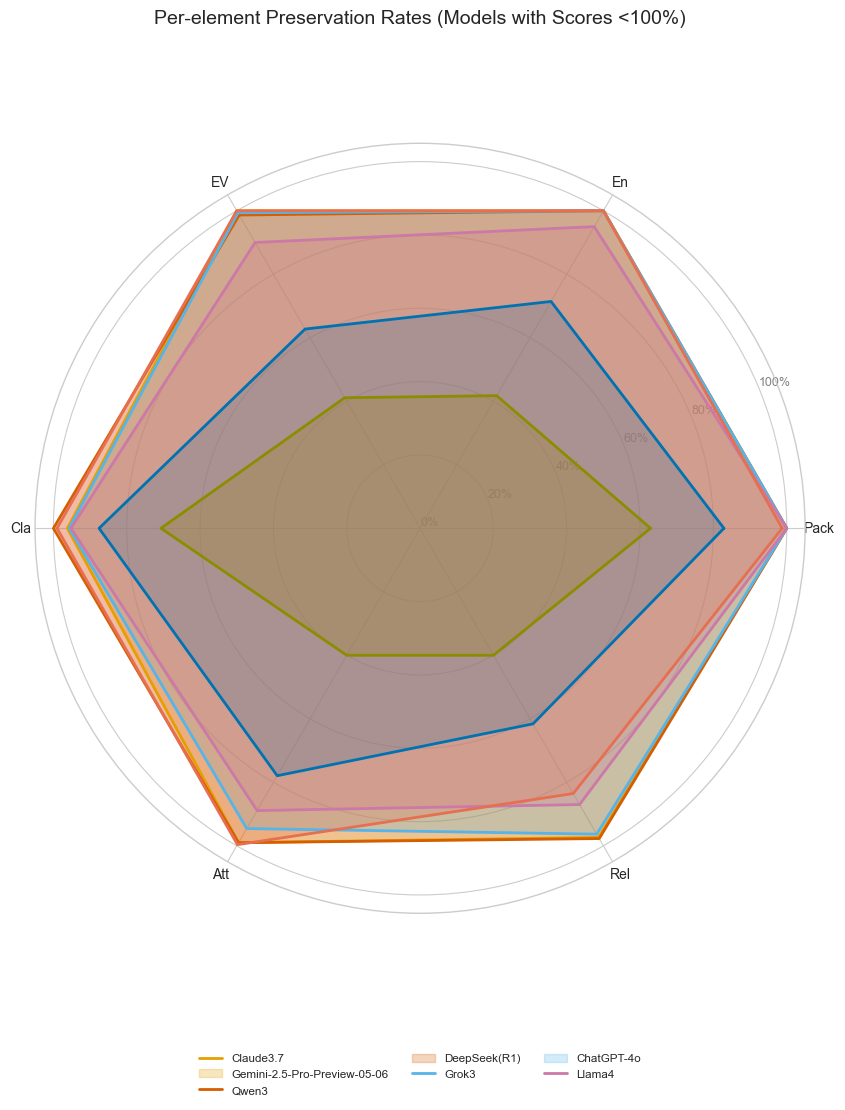}
  \caption{Preservation rates by UML element for models with subperfect performance.}
  \label{fig:sf-radar}
\end{figure}

Structural fidelity, measured by Jaccard similarity across six UML element categories, is summarized in Table~\ref{tab:sf-detail} and visualized in Figures~\ref{fig:sf-bar} and \ref{fig:sf-radar}.

\begin{table}[ht]
  \centering
  \caption{Mean preservation rates for UML structural elements by model.}

  \label{tab:sf-detail}
  \small
  \begin{tabular}{lccccccc}
    \toprule
    \textbf{Model} & \textbf{Pkg} & \textbf{Enum} & \textbf{Ev} & \textbf{Cls} & \textbf{Attr} & \textbf{Rel} & \textbf{Glob} \\
    \midrule
    Claude~3.7       & 100.00 & 100.00 & 100.00 & 96.19 & 99.02 & 97.33 & 98.76 \\
    Gemini~2.5~Pro   & 100.00 & 100.00 & 98.65  & 100.00 & 99.02 & 97.71 & 99.23 \\
    ChatGPT-o3       & 100.00 & 100.00 & 100.00 & 100.00 & 100.00 & 100.00 & 100.00 \\
    Qwen~3           & 100.00 & 100.00 & 99.35  & 95.71  & 94.56 & 96.33  & 97.66 \\
    DeepSeek~R1      & 100.00 & 95.00  & 90.00  & 95.21  & 88.90 & 87.00  & 92.69 \\
    Grok~3           & 82.86  & 71.43  & 62.74  & 87.55  & 77.92 & 61.59  & 74.01 \\
    ChatGPT-4o       & 62.86  & 41.76  & 41.11  & 70.67  & 40.00 & 40.00  & 49.40 \\
    Llama~4          & 98.75  & 100.00 & 100.00 & 99.02  & 99.63 & 83.54  & 96.82 \\
    Mistral          & 100.00 & 100.00 & 100.00 & 100.00 & 100.00 & 100.00 & 100.00 \\
    \bottomrule
  \end{tabular}
\end{table}

Mixed ANOVA results indicate significant main effects of model (\(F(8,81) = 7.78, p < .0001, \eta^2_G = 0.380\)) and UML element type (\(F(5,405) = 9.55, p < .0001, \eta^2_G = 0.023\)), as well as a significant model-by-element interaction (\(F(40,405) = 2.87, p < .0001, \eta^2_G = 0.054\)).

Models including \textbf{ChatGPT-o3}, \textbf{Mistral}, \textbf{Gemini~2.5~Pro}, \textbf{Claude~3.7}, and \textbf{Llama~4} preserve the baseline UML structure near perfectly. In contrast, \textbf{Grok~3} and \textbf{ChatGPT-4o} show diminished preservation, particularly for enumerations and relationships, suggesting occasional omissions or modifications.

\paragraph{Interpretation and Practical Implications}  
Maintaining high structural fidelity is critical in regulated and safety-critical software domains where preserving UML semantics supports compliance, reliability, and risk mitigation throughout development and maintenance. High fidelity also facilitates seamless integration of behavioral enrichments in collaborative workflows. Conversely, lower fidelity may introduce inconsistencies complicating downstream tasks. Therefore, selecting LLMs with strong structural preservation is essential for balancing accuracy and robustness in software engineering practice.

\subsection{Syntactic Correctness (SC)}

Syntactic correctness was evaluated based on PlantUML compilation success. Table~\ref{tab:sc-contingency} summarizes the counts of successfully rendered versus errored diagrams for each LLM.

\begin{table}[ht]
  \centering
  \caption{Compilation results for PlantUML diagrams.}

  \label{tab:sc-contingency}
  \small
  \begin{tabular}{lcc}
    \toprule
    \textbf{LLM} & \textbf{Correct Diagrams} & \textbf{Errored Diagrams} \\
    \midrule
    Claude~3.7      & 10 & 0 \\
    Gemini~2.5~Pro  & 8  & 2 \\
    ChatGPT-o3      & 10 & 0 \\
    Qwen~3          & 10 & 0 \\
    DeepSeek~R1     & 9  & 1 \\
    Grok~3          & 9  & 1 \\
    ChatGPT-4o      & 9  & 1 \\
    Llama~4         & 9  & 1 \\
    Mistral         & 10 & 0 \\
    \bottomrule
  \end{tabular}
\end{table}

No statistically significant differences in syntactic correctness rates were observed across models (\(\chi^2(8) = 6.43, p = 0.5993\)), despite a moderate effect size (Cramér’s \(V = 0.267\)). The uniformly high syntactic correctness confirms these LLMs’ outputs are immediately usable within UML toolchains.

\paragraph{Interpretation and Practical Implications}  
The consistently high syntactic validity across all models provides a strong foundation for integrating LLM-generated UML enrichments into existing modeling workflows with minimal manual intervention, thereby accelerating development cycles and reducing overhead.

\subsection{Method Name Convergence}

This subsection analyzes lexical and structural consensus among LLM-generated methods through Top-Method Consensus (TMC), Core-Method Consensus (CMC), and Structural Placement Consistency (SPC).
\subsubsection{Top-Method Consensus (TMC)}

\begin{figure}[ht]
  \centering
  \includegraphics[width=\linewidth]{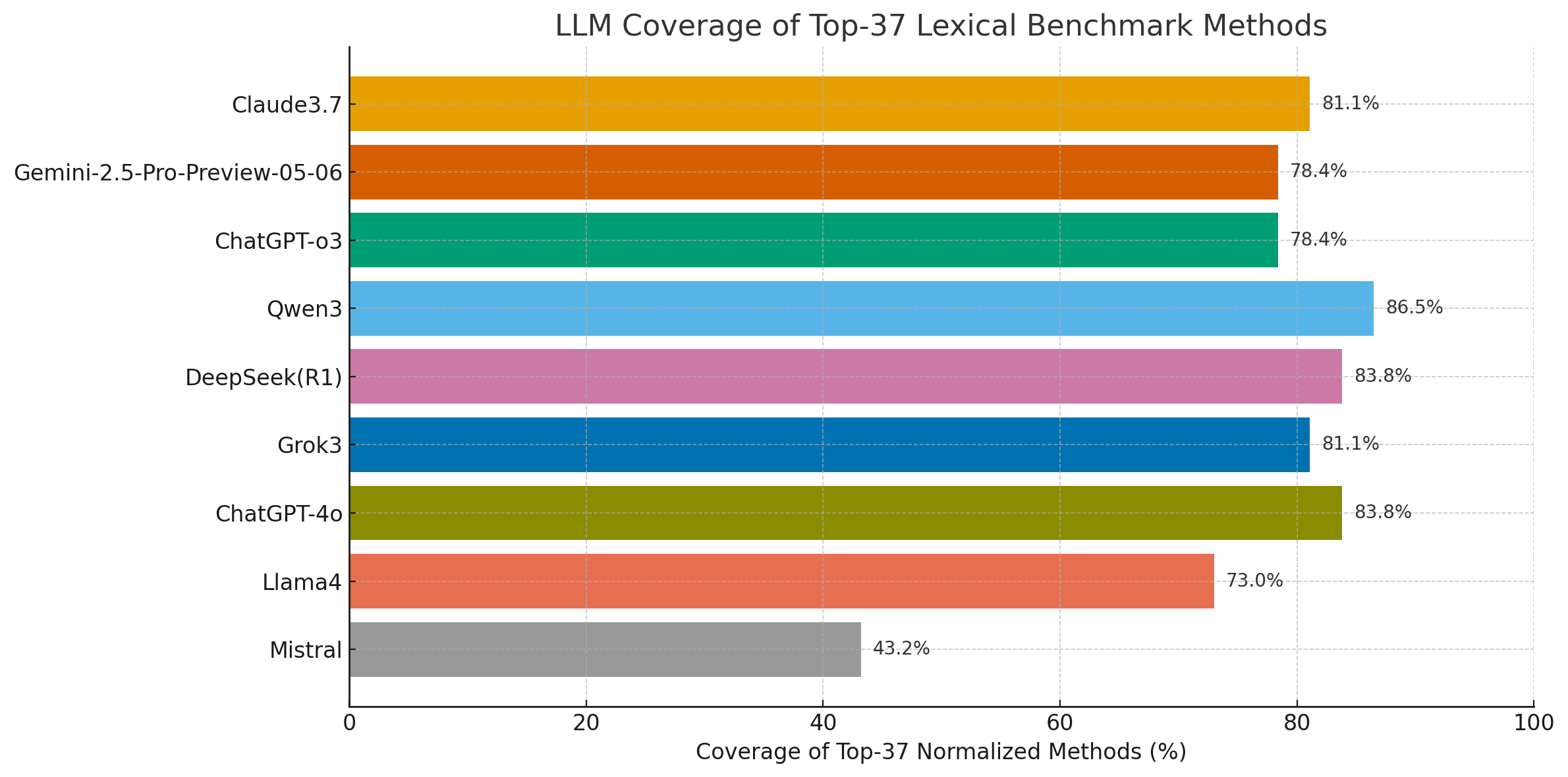}
  \caption{Coverage of Top-37 normalized benchmark methods by each LLM.}
  \label{fig:tmc-bar-coverage}
\end{figure}

The Top-37 most frequent normalized method names—where normalization involves converting method names to a canonical form by lowercasing and removing formatting differences such as underscores and camelCase distinctions—were extracted as a benchmark set. Coverage by each LLM varies significantly (Figure~\ref{fig:tmc-bar-coverage}), ranging from 43.2\% (Mistral) to 86.5\% (Qwen~3). A Kruskal–Wallis test confirms this difference is statistically significant (H=26.42,p=0.0009H=26.42,p=0.0009) with a moderate effect size.

\begin{figure}[ht]
  \centering
  \includegraphics[width=\linewidth]{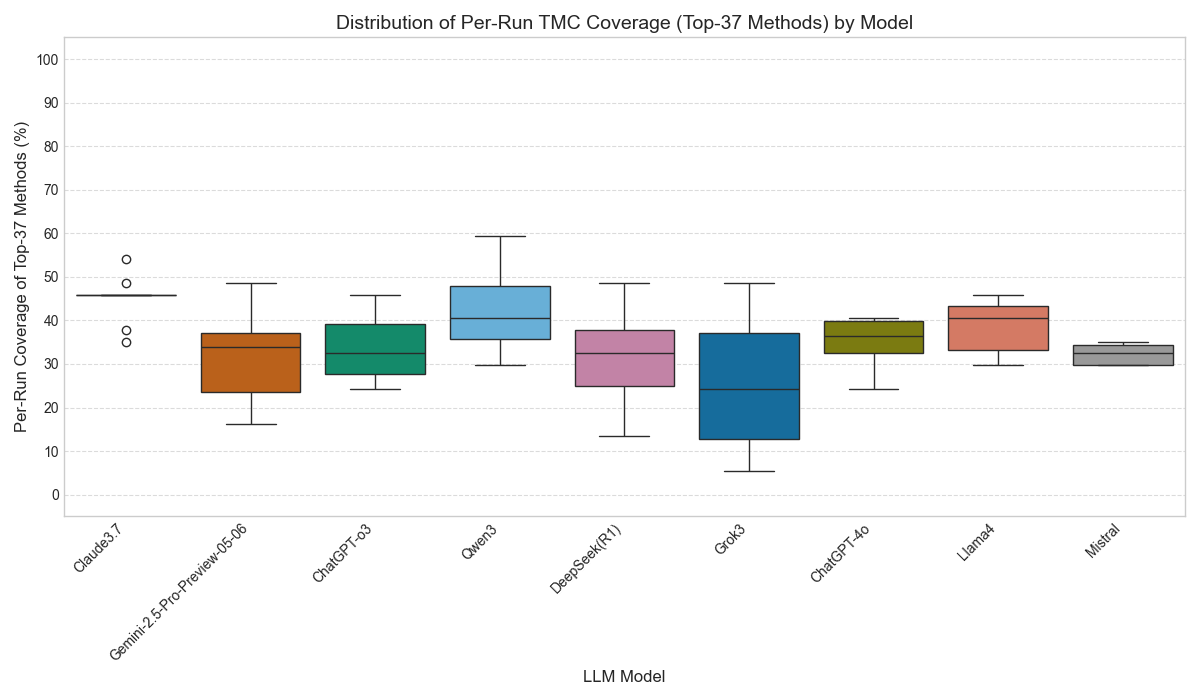}
  \caption{Distribution of per-run TMC coverage by model.}
  \label{fig:tmc-boxplot}
\end{figure}

Examining per-run distributions reveals variability in consensus (Figure~\ref{fig:tmc-boxplot}). \textbf{Claude~3.7} achieves both high mean coverage and low variance, indicating stable reproduction of core methods across runs. In contrast, models such as \textbf{Grok~3} and \textbf{DeepSeek~R1} show wider spreads and lower medians, suggesting inconsistent capture of these key methods.

\begin{figure}[ht]
  \centering
  \includegraphics[width=\linewidth]{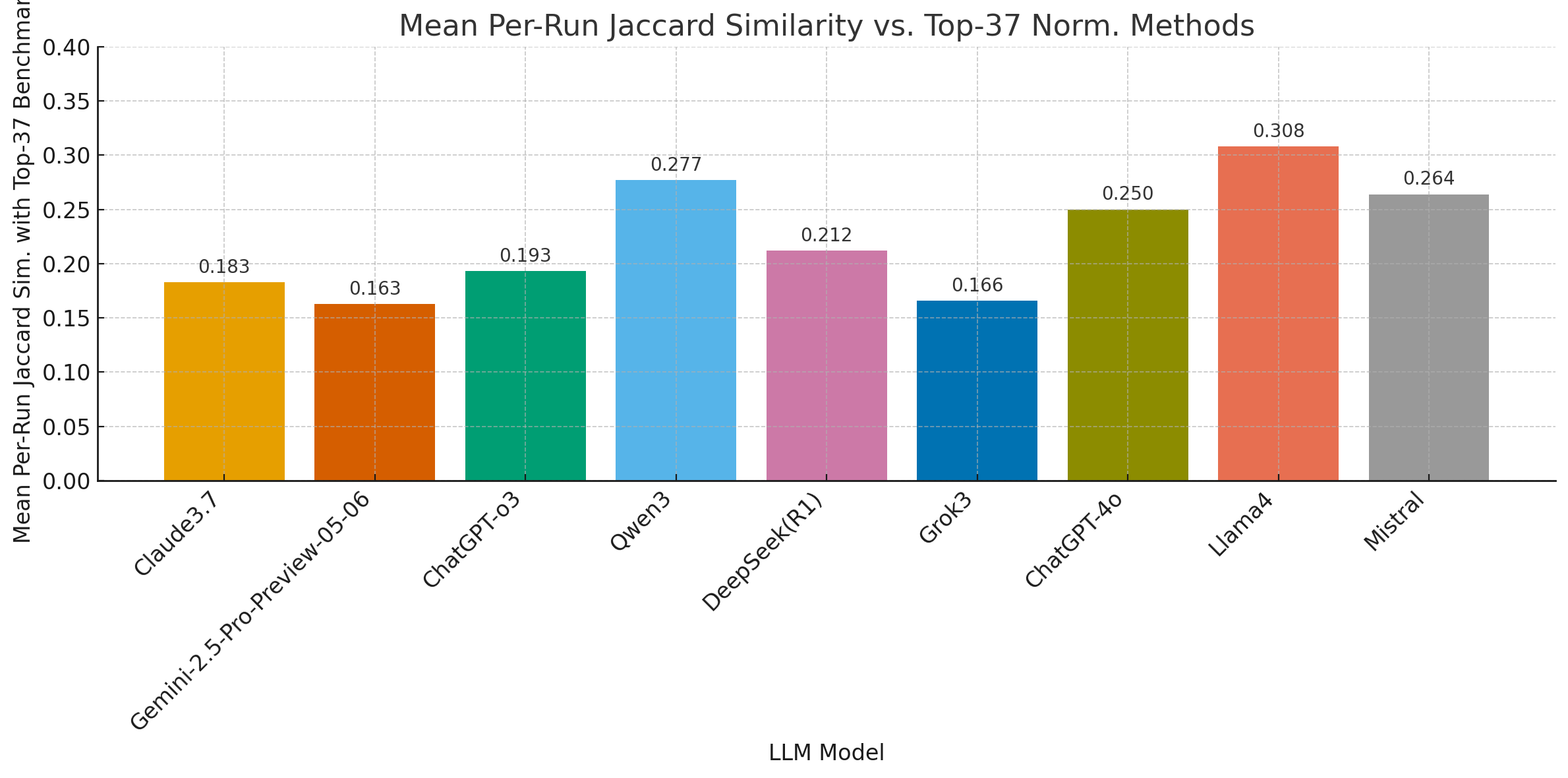}
  \caption{Mean per-run Jaccard similarity to Top-37 normalized benchmark.}
  \label{fig:tmc-jaccard-bar}
\end{figure}

Jaccard similarity to the Top-37 benchmark also varies significantly among models (\(H = 45.72, p < 0.0001\); Figure~\ref{fig:tmc-jaccard-bar}), highlighting differences in the specific method subsets reproduced, even among top performers.

\paragraph{Interpretation and Practical Implications}  
Top-method consensus reflects the degree to which LLMs reproduce a common set of high-frequency, normalized method names, serving as a proxy for lexical convergence and alignment with typical behavioral operations. High coverage and stability (e.g., Qwen~3, Claude~3.7) indicate strong consistency beneficial for collaborative modeling or multi-agent generation scenarios. Conversely, lower and more variable coverage (e.g., Grok~3) may lead to fragmented behavior definitions, necessitating manual harmonization. Per-run stability further differentiates models suited for applications requiring reliable and predictable behavioral scaffolding.
\subsubsection{Core-Method Consensus (CMC)}

\begin{figure}[ht]
\centering
\includegraphics[width=\linewidth]{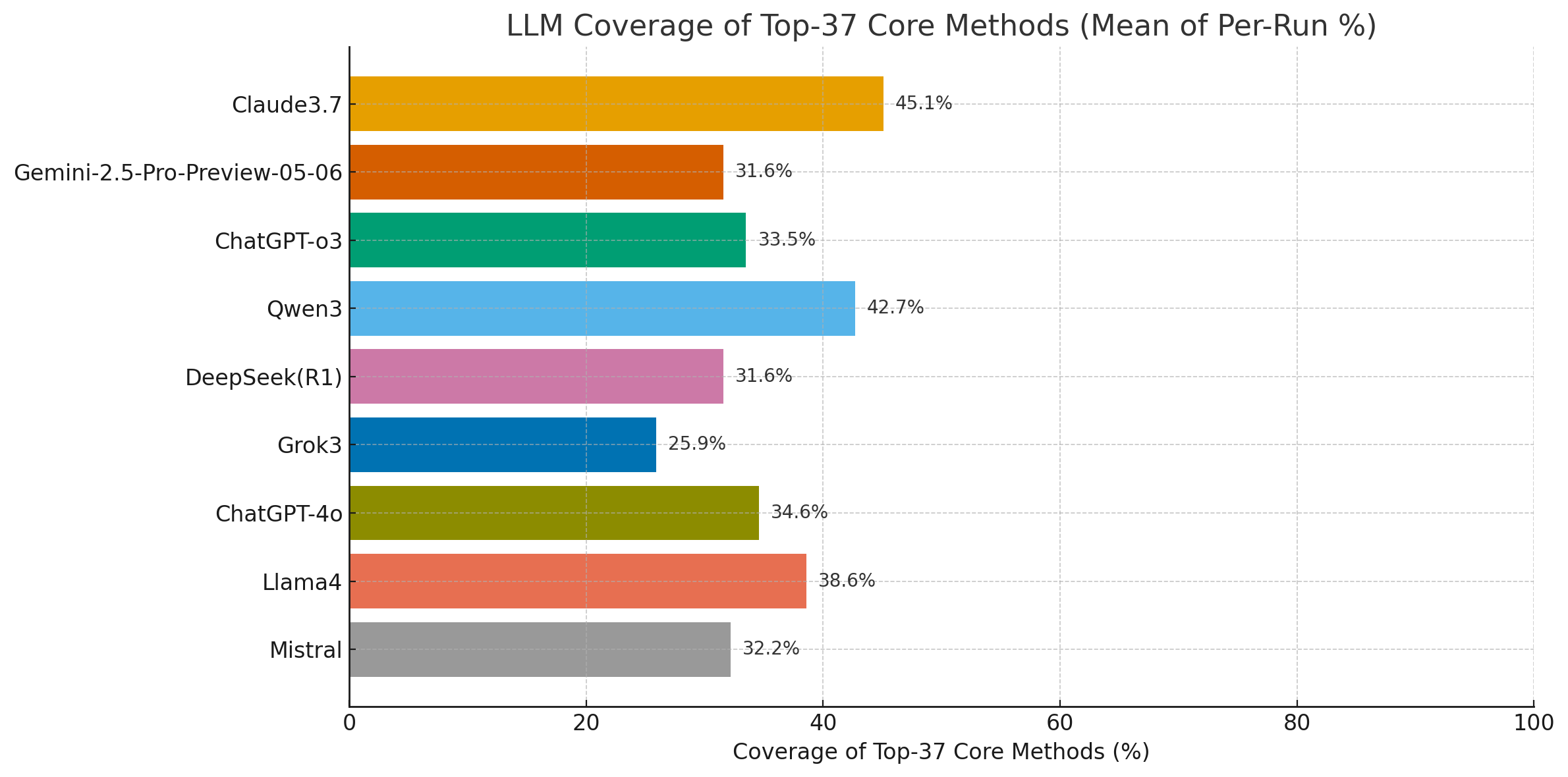}
\caption{Coverage of top-37 core methods by LLM (mean per-run percentage).}
\label{fig:CC-bar}
\end{figure}

Coverage of the 37 most frequently generated raw method names serves as a proxy for lexical consensus across LLMs. Figure~\ref{fig:CC-bar} summarizes the average per-run coverage of these core methods by each model. \textbf{Claude~3.7} and \textbf{Qwen~3} lead with 45.1\% and 42.7\% coverage, respectively, while \textbf{Grok~3} trails at 25.9\%. Kruskal–Wallis tests confirm significant differences across models (\(H = 26.42, p = 0.0009\)).

\begin{figure}[ht]
\centering
\includegraphics[width=0.95\linewidth]{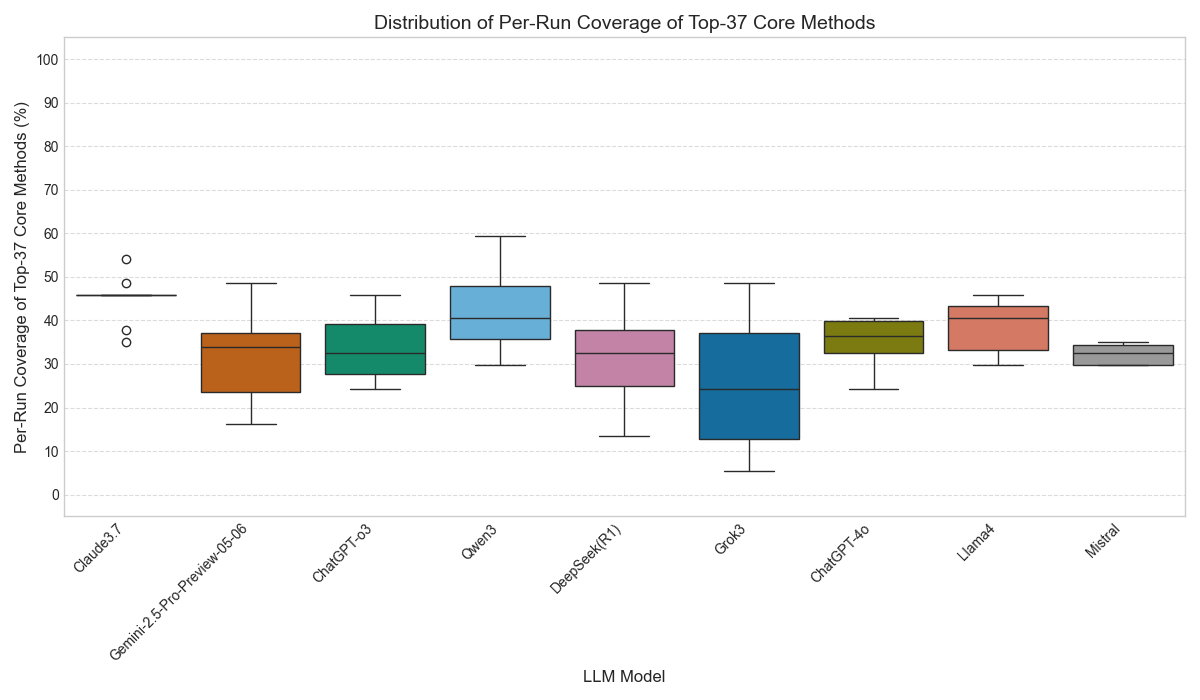}
\caption{Distribution of per-run core-method coverage by model.}
\label{fig:CC-boxplot}
\end{figure}

To evaluate variability and consistency across runs, Figure~\ref{fig:CC-boxplot} shows the per-run distribution of core-method coverage per model. Notably, \textbf{Claude~3.7} exhibits both high median coverage and low dispersion, indicating stable reproduction of core methods. In contrast, \textbf{Grok~3} and \textbf{DeepSeek~R1} display wider variance and lower medians, suggesting less consistent capture of the shared method set.

\begin{figure}[ht]
\centering
\includegraphics[width=0.6\linewidth]{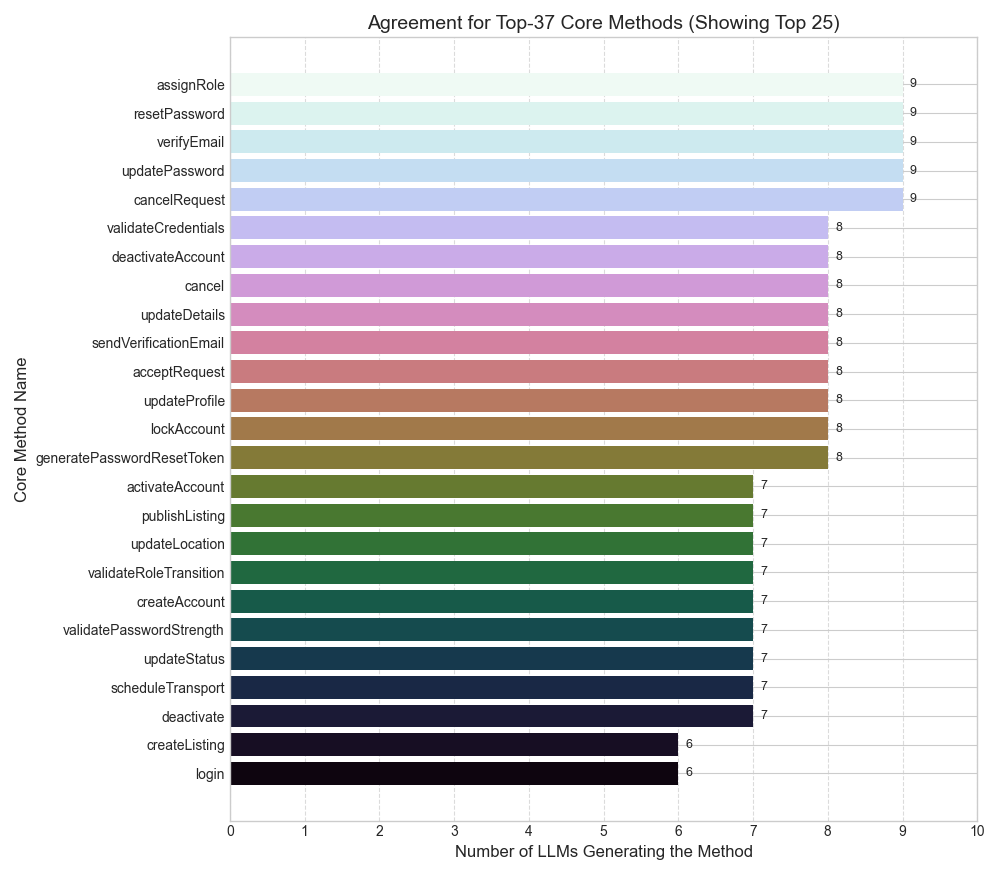}
\caption{Model agreement per core method: number of LLMs generating each of the top-37 core methods.}
\label{fig:CC-agreement}
\end{figure}

Figure~\ref{fig:CC-agreement} visualizes model agreement for individual core methods. Methods such as \texttt{assignRole}, \texttt{verifyEmail}, \texttt{updatePassword}, and \texttt{cancelRequest} were generated by all nine models, reflecting behaviors that are semantically directly tied to corresponding actions within main scenario use cases (UCs). This indicates these operations are universally recognized as fundamental domain behaviors. Mid-ranked methods (e.g., \texttt{publishListing}, \texttt{scheduleTransport}) show moderate agreement, while tail-end methods exhibit greater divergence.

\begin{figure}[ht]
\centering
\includegraphics[width=0.5\linewidth]{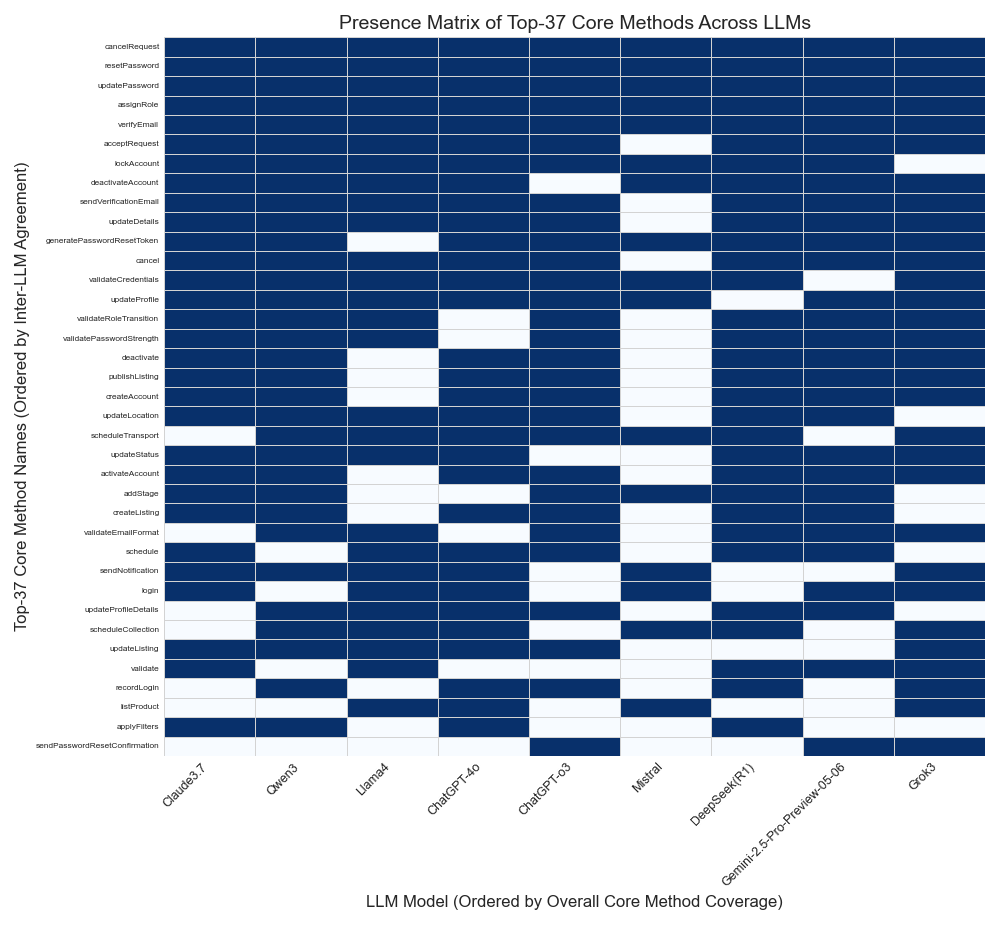}
\caption{Presence matrix of top-37 core methods across LLMs; dark cells denote inclusion.}
\label{fig:CC-heatmap}
\end{figure}

Complementing this, Figure~\ref{fig:CC-heatmap} presents a binary presence matrix of core methods across models. The clustered arrangement reveals high-agreement clusters (e.g., top-left block) alongside fragmented areas, highlighting which models consistently contribute to consensus and which omit specific methods.

\paragraph{Interpretation and Practical Implications}
Core-method consensus highlights which operations are most frequently and consistently identified across LLMs, reflecting a shared behavioral understanding grounded in key domain actions. The methods generated by all nine models correspond directly to critical actions within the main scenario use cases, underscoring their fundamental role in domain modeling. High-performing models like \textbf{Claude3.7} and \textbf{Qwen3} demonstrate strong convergence on these domain-relevant behaviors, suggesting greater reliability and suitability for consensus-driven or ensemble-based design workflows. Conversely, lower agreement from models such as \textbf{Grok~3} may indicate broader generative variability or misalignment with core behavioral norms. These insights can inform model selection, support automated quality checks in AI-assisted tooling, and help establish minimum behavioral standards for expert validation.

\subsubsection{Structural Placement Consistency (SPC)}

Structural Placement Consistency (SPC) quantifies how consistently core methods are assigned to the same UML classes across different LLMs. The average placement consistency is high at 92.7\%, with a Wilcoxon signed-rank test revealing a slight but statistically significant deviation from perfect agreement (\(p = 0.0112\)).

Such minor deviations from perfect consistency are expected in complex modeling tasks and likely reflect valid alternative architectural interpretations rather than errors.

\begin{figure}[ht]
  \centering
  \includegraphics[width=0.85\linewidth]{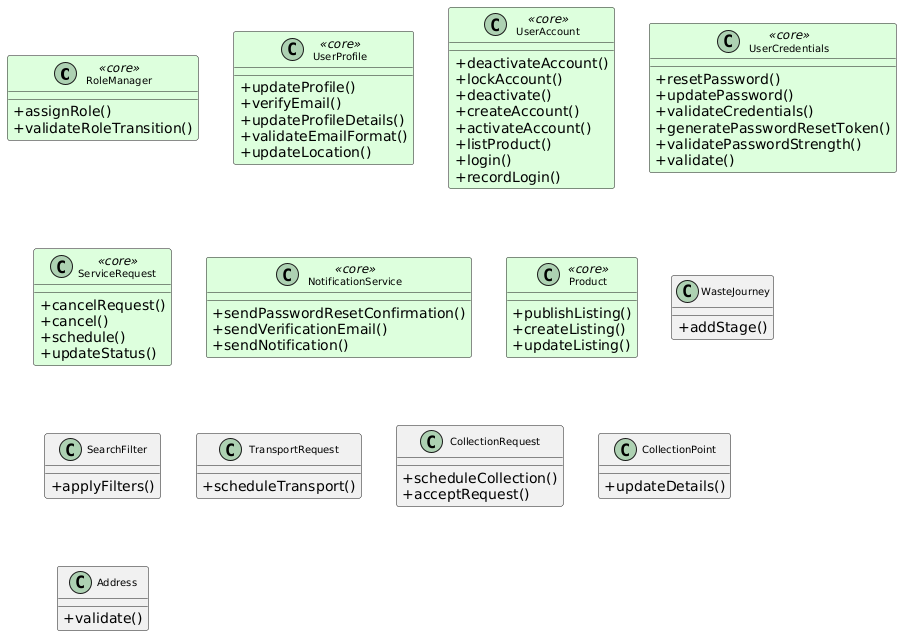}
  \caption{UML diagram of 37 core methods grouped by dominant class; green classes show 100\% placement consistency.}
  \label{fig:core-methods-full-uml}
\end{figure}

Table~\ref{tab:spc-core-methods} (Appendix~\ref{appendix:spc-core-methods}) provides detailed placement consistency scores for each core method, while Figure~\ref{fig:core-methods-full-uml} visualizes core methods grouped by their dominant class.

\paragraph{Interpretation and Practical Implications}  
The high SPC observed indicates strong convergence among LLMs not only in method naming but also in architectural reasoning, suggesting that models reliably infer appropriate class-method associations from use-case semantics. Figure~\ref{fig:core-methods-full-uml} bridges quantitative evaluation with planned semantic validation, visually depicting core methods grouped by dominant class—many achieving perfect placement consistency (highlighted in green). This underscores meaningful method distribution and architectural coherence, which is critical in collaborative or multi-model design environments to reduce integration conflicts and support maintainability. Minor inconsistencies point to valid alternative interpretations, highlighting the need for enhanced semantic disambiguation in prompt design and validation pipelines. These implications and future research directions are discussed in detail in the Discussion section.

\subsection{Summary}  
\label{subsec:summary}

This study evaluates the ability of large language models (LLMs) to enrich UML class diagrams with methods derived from structured use cases. The key findings are:

\begin{itemize}
    \item \textbf{Method Quantity (MQ)}: \textbf{Claude~3.7} generated the highest number of methods, providing comprehensive behavioral coverage, while models like \textbf{Mistral} and \textbf{Llama~4} produced fewer methods, favoring simplicity.
    \item \textbf{Signature Richness (SR)}: \textbf{Mistral} produced the richest method signatures, particularly excelling in parameter richness and return type annotations, whereas other models showed more conservative parameterization.
    \item \textbf{Annotation Completeness (AC)}: \textbf{Claude~3.7} and \textbf{ChatGPT-o3} had the highest annotation adherence, ensuring traceability from use cases to methods, which is essential for maintainability and collaboration.
    \item \textbf{Structural Fidelity (SF)}: \textbf{ChatGPT-o3}, \textbf{Mistral}, and \textbf{Claude~3.7} preserved the UML diagram structure with high consistency, while models like \textbf{Grok~3} showed weaker structural preservation, particularly for relationships and enumerations.
    \item \textbf{Novelty - Method Name Consensus}: A notable finding is the high lexical consensus in method naming across models. \textbf{Claude~3.7} and \textbf{Qwen~3} demonstrated strong consistency in method names, which is critical for maintaining uniformity across multiple LLMs in collaborative workflows.
    \item \textbf{Novelty - Multi-model Evaluation}: This study's multi-model approach is a significant contribution, providing a broader view of LLM performance. Unlike previous work that focused on single models, we offer a comprehensive benchmark, showing how different LLMs perform across multiple dimensions.
\end{itemize}

In conclusion, while LLMs show promise for automating UML behavioral enrichment, model selection should be based on specific project needs such as method quantity, signature detail, and annotation completeness. The high degree of consensus across models highlights the potential for cross-model collaboration, which could enable more robust and consistent UML design processes in software engineering.

\section{Discussion}
\label{sec:discussion}

This study presents a systematic evaluation of nine publicly available state-of-the-art large language models (LLMs) in their ability to automatically enrich UML class diagrams with methods derived from 21 structured natural language use cases. The central research question addressed is:

\begin{quote}
\emph{How effectively do nine publicly available LLMs enrich structural UML class diagrams with methods derived from structured use cases?}
\end{quote}

By analyzing multiple quantitative metrics across 3,373 generated methods spanning 90 UML diagrams, an empirical foundation is established that characterizes the capabilities, limitations, and practical considerations for deploying LLMs in automated UML enrichment.

In addition to effectiveness, the practical efficiency of the LLM-based enrichment process was notable. Across all runs, each LLM required between a few seconds and under five minutes to analyze the input artifacts—including the methodless UML class diagram, the structured use cases, and enrichment instructions—and generate the fully updated PlantUML code. This rapid end-to-end processing substantially exceeds typical human manual efforts, making it well-suited to Agile software development practices that emphasize fast iteration, continuous feedback, and incremental design refinement. The capability to quickly produce and update UML behavioral models aligns with Agile’s need for adaptive modeling and supports more responsive and collaborative engineering workflows.

\subsection{Summary of Key Metrics}

To clarify the evaluation framework, model outputs were assessed along six key dimensions:

\begin{itemize}
    \item \textbf{Method Quantity (MQ):} Number of methods generated per UML diagram.
    \item \textbf{Signature Richness (SR):} Detail and complexity of method parameters and lexical diversity of method names.
    \item \textbf{Annotation Completeness (AC):} Presence of source-use-case and action references linking methods to their originating requirements.
    \item \textbf{Structural Fidelity (SF):} Preservation of original UML class diagram relationships and structure.
    \item \textbf{Syntactic Correctness (SC):} Validity of generated PlantUML code based on successful compilation.
    \item \textbf{Method Name Consensus (MNC):} Agreement among models in naming conventions for methods reflecting shared domain understanding.
\end{itemize}

These metrics collectively characterize the quantitative aspect of the enriched UML diagrams.

\subsection{Model Performance Highlights and Selection Considerations}

Multi-dimensional analysis reveals distinct strengths and trade-offs among the LLMs:

\paragraph{Method Quantity (MQ)}  
Claude~3.7 generated the largest number of methods (734), significantly outperforming others such as Gemini~2.5Pro (477) and ChatGPT-o3 (432). In contrast, models like Mistral8$\times$7B and Llama~4 adopted more conservative generation strategies. Statistical tests (\textit{Kruskal-Wallis} $H(8)=51.85$, $p<.0001$) confirm these differences. Practitioners should align model choice with project needs: Claude~3.7 and ChatGPT-o3 are suitable for comprehensive coverage, while leaner models may better support maintainability.

\paragraph{Signature Richness (SR)}  
Signature richness was evaluated across five dimensions: visibility markers, naming conventions, parameter richness, return types, and lexical diversity. All models predominantly used the public visibility marker (`+`), with a statistically significant but small variation in visibility usage across models (\(\chi^2 = 64.09, p < .0001\)). Naming conventions were uniformly camelCase (100\%), demonstrating adherence to industry standards.

Parameter richness varied significantly among models (\(H = 382.08, p < .0001\)), with Mistral generating notably richer parameter lists than others, supported by large effect sizes (Cliff’s \(\delta = 1.0\) versus Claude~3.7). Return-type annotations also differed significantly (\(\chi^2 = 312.38, p < .0001\)), with Mistral and Claude~3.7 exhibiting higher proportions of non-void return types, indicating enhanced behavioral detail.

Lexical diversity, measured via normalized Levenshtein distance among unique method names, varied significantly (\(H = 44.07, p < .0001\)), highlighting differences in naming expressiveness and redundancy. Models such as Claude~3.7 and ChatGPT variants exhibited higher lexical diversity, producing more descriptive and varied method names, while others like Grok~3 favored more conservative naming with less variation.

These variations in signature richness reflect differing model design philosophies: models like Mistral prioritize detailed behavioral signatures at the cost of potential complexity, whereas models favoring lexical diversity generate clearer and more maintainable method names. This highlights the importance of aligning LLM selection with project priorities that balance parameter detail and naming clarity.

\paragraph{Annotation Completeness (AC)}  
Traceability through annotations linking methods to their originating use cases and actions was strongest in Claude~3.7 and ChatGPT-o3. Conversely, Mistral8$\times$7B and Llama~4 often omitted such annotations, which may reduce interpretability and hinder downstream validation. This relationship was statistically significant ($\chi^2(16)=3001.07$, $p<.0001$), highlighting annotation discipline as a key differentiator among models.

\paragraph{Structural Fidelity (SF)}  
ChatGPT-o3, Mistral8$\times$7B, and Claude~3.7 preserved over 98\% of the original UML class relationships, maintaining high structural consistency. Other models, such as Grok3 and ChatGPT-4o, showed weaker preservation. Mixed ANOVA results confirmed significant effects of model choice on structural fidelity ($F(8,81)=7.78$, $p<.0001$). This suggests some LLMs better maintain diagram integrity than others.

\paragraph{Syntactic Correctness (SC)}  
All evaluated LLMs achieved greater than 90\% success rates in generating PlantUML code that compiled without errors, indicating reliable syntactic validity across models. No significant differences were detected (\(\chi^2(8)=6.43\), \(p=0.599\)), supporting the feasibility of toolchain integration regardless of model selection.

Despite uniformly high syntactic correctness, a moderate effect size (Cramér’s \(V = 0.267\)) was observed, suggesting subtle model-specific differences that may warrant attention in particular integration scenarios, even though overall differences were not statistically significant.

\paragraph{Method Name Consensus (MNC)}  
Models Qwen3, Claude~3.7, and ChatGPT-o3 demonstrated higher agreement in method naming conventions, reflecting shared domain understanding. Mistral8$\times$7B and Grok3 exhibited more divergent naming patterns. These differences were statistically significant (\textit{Kruskal-Wallis} $H(8)=26.42$, $p=0.0009$), indicating variation in lexical consistency across models.

\subsubsection*{Model Selection Guidelines}

Based on these findings, the following recommendations are proposed:

\begin{itemize}
    \item \textbf{For comprehensive coverage and completeness:} Claude~3.7 or ChatGPT-o3
    \item \textbf{For detailed parameterization:} Mistral8$\times$7B, recognizing a trade-off with annotation completeness
    \item \textbf{For balanced output:} Gemini~2.5Pro
\end{itemize}

For mission-critical applications, ensemble approaches that combine complementary model strengths may offer superior performance.

\subsection{Method Name Consensus and Structural Placement Consistency: Key Novel Insight}

A striking and unexpected finding is the high degree of consensus exhibited among diverse LLMs in generating UML method names and structurally assigning them to classes. Despite differences in generation approaches and parameter richness, models consistently produced a core set of method names with substantial overlap.

For example, methods universally agreed upon by all models—such as \texttt{assignRole}, \texttt{verifyEmail}, \texttt{updatePassword}, and \texttt{cancelRequest}—correspond directly to fundamental actions within key use cases. This highlights that LLMs internalize essential domain semantics, enabling semantically coherent and architecturally consistent behavioral enrichments.

Coverage of the top 37 normalized method names ranged from 43.2\% (Mistral) to 86.5\% (Qwen~3), with several models exceeding 80\% coverage. This lexical agreement is complemented by a high degree of structural placement consistency, with an average of 92.7\% of core methods assigned to the same UML classes across models.

Statistical analysis of Structural Placement Consistency (SPC) revealed a slight but significant deviation from perfect agreement among models (p=0.0112, Wilcoxon signed-rank test), indicating minor variations in method placement likely reflecting valid alternative interpretations rather than errors.

This convergence, observed without explicit shared prompts or references, provides a novel empirical basis for enhancing trust in AI-assisted UML design.

Practically, this supports ensemble or consensus-based approaches, where methods and placements agreed upon by multiple models can be prioritized for automation and expert validation, improving efficiency while maintaining quality.

From a research standpoint, these results invite further exploration into the mechanisms underlying consensus formation among LLMs and offer a promising proxy for semantic validation in the absence of ground truth benchmarks.

This strong consensus represents a key contribution of this investigation, advancing both applied practice and foundational understanding of AI-driven software engineering.

\subsection{Interpretation and Practical Implications}

These results extend prior AI-assisted design research by quantitatively evaluating LLM-based UML behavioral enrichment across multiple dimensions. Key implications include:

\begin{enumerate}
    \item \textbf{Capability-Reliability Gap:} Although LLMs reliably generate syntactically valid and structurally consistent UML enrichments, considerable variability remains in method quantity, signature detail, annotation quality, and structural preservation. Notably, syntactic or structural correctness does not guarantee semantic validity or alignment with intended functionality.
    
    \item \textbf{Automation Potential:} Consistent naming and placement consensus among models suggests LLMs capture core domain behaviors effectively, enabling automation of routine enrichment tasks to accelerate design workflows.
    
    \item \textbf{Human Oversight Necessity:} Variability in annotation completeness and signature richness underscores the need for expert review to ensure traceability, interpretability, and functional correctness.
    
    \item \textbf{Collaborative Human-AI Workflow:} Treating LLMs as collaborative partners that automate initial design drafts, with experts providing semantic validation, refinement, and contextual judgment, is recommended.
    \item \textbf{Agile-friendly practices:} The ability of LLMs to rapidly generate and update UML enrichments supports Agile methodologies by facilitating iterative design, rapid prototyping, and adaptive workflows aligned with evolving project needs.

\end{enumerate}

\subsection{Limitations}

\paragraph{Semantic Validation}  
Open-ended method generation precludes definitive automated benchmarks. This study relies on proxy metrics rather than direct verification of behavioral alignment, method-to-class assignment accuracy, or comprehensive semantic coverage.

\paragraph{Benchmark Scarcity}  
The lack of publicly available datasets linking use cases to fully behaviorally complete UML diagrams limits rigorous semantic evaluation.

\paragraph{Operational Constraints}  
LLM outputs are sensitive to input quality and fixed prompting strategies, and the probabilistic nature of LLMs necessitates human oversight due to variability.

Furthermore, the fixed prompt structure and reliance on single-shot generation may limit output diversity and adaptability. Variability in LLM API versions and environmental factors can influence reproducibility, highlighting the need for standardized prompt engineering and monitoring in operational deployments.

\paragraph{Domain Specificity}  
Evaluation focused on the waste management domain, which may limit applicability to safety-critical or other specialized systems.

\subsection{AI Bias Considerations}

Observed output variations may reflect underlying biases present in training corpora, risking perpetuation of representational limitations. Effective mitigation requires development of domain-diverse datasets, algorithmic bias detection techniques, and rigorous validation processes. Future research should quantify bias manifestations and develop domain-adapted debiasing frameworks.

\subsection{Threats to Validity}

\begin{itemize}
    \item \textbf{Construct validity:} Proxy metrics do not directly capture semantic or behavioral correctness.
    \item \textbf{Internal validity:} Fixed prompt sequences and single-domain datasets may bias model outputs. Multiple random seeds were used to mitigate stochastic effects, though API changes may impact reproducibility.
    \item \textbf{External validity:} Domain and UML diagram type specificity limit generalizability. Only publicly accessible LLMs available during the study period were evaluated.
    \item \textbf{Reliability:} Results depend on consistent model access and prompt configuration; environmental variability may influence output consistency.
    \item \textbf{Statistical validity:} Appropriate statistical tests and corrections for multiple comparisons were applied, though inherent LLM output variability remains a confounding factor.
\end{itemize}

Expanding evaluation domains, integrating semantic validation techniques, and conducting longitudinal studies are necessary to mitigate these threats.

\subsection{Future Research Directions}

Several promising avenues for continued investigation are identified:

\begin{enumerate}
    \item \textbf{Semantic Validation:} Development of integrated frameworks combining expert review, formal verification, and behavioral testing.
    \item \textbf{Benchmark Creation:} Establishment of expert-validated datasets linking natural language specifications to behaviorally complete UML models.
    \item \textbf{Consensus for Semantics:} Further work should investigate how consensus measures can inform and complement semantic validation approaches.
    \item \textbf{Model Enhancement:} Exploration of adaptive prompting and domain-specific fine-tuning to improve output relevance.
    \item \textbf{Hybrid Architectures:} Investigation of ensemble methods leveraging complementary strengths of diverse LLMs.
    \item \textbf{Generalizability Studies:} Extension of evaluations to multiple domains, including safety-critical systems.
    \item \textbf{Human-AI Collaboration Workflows:} Design of interactive refinement protocols integrating human expertise with AI assistance.
    \item \textbf{Bias Mitigation:} Development of domain-specific debiasing techniques and bias detection tools.
    \item \textbf{Validation Pipelines:} Creation of integrated metric-feedback systems to continuously monitor and improve output quality.
\end{enumerate}

Bridging syntactic-structural evaluation with semantic validation and domain expertise remains critical to advancing trustworthy AI-assisted modeling.

\subsection{Reflections on Human-AI Collaboration}

Findings reaffirm that while LLMs generate syntactically valid UML artifacts proficiently, critical human oversight remains indispensable to ensure accuracy, appropriateness, and semantic alignment—particularly given input ambiguities and probabilistic variability. A collaborative paradigm is proposed in which LLMs automate repetitive or draft generation tasks, and experts provide domain knowledge, semantic validation, and refinement. This division balances efficiency gains with nuanced judgment.

The collaborative paradigm between humans and LLMs aligns closely with Agile software development principles, emphasizing iterative refinement, continuous feedback, and adaptive workflows. By automating routine and repetitive tasks, LLMs enable faster generation of initial UML design drafts, allowing experts to focus on semantic validation and nuanced decision-making. This division of labor supports the Agile values of collaboration and responsiveness, facilitating rapid iteration cycles and evolving models that better reflect changing requirements. Consequently, integrating LLM-assisted enrichment within Agile workflows can enhance both efficiency and design quality through synergistic human-AI interaction.

\subsection{Concluding Remarks}

This multi-dimensional evaluation of nine publicly accessible LLMs reveals significant variability in their ability to enrich UML class diagrams with methods derived from structured use cases. While syntactic robustness and structural fidelity are generally assured, challenges remain in semantic validation, bias mitigation, and consistency. Context-driven model selection and ensemble strategies show promise for practical adoption. This work establishes a rigorous empirical foundation supporting trustworthy integration of LLMs into UML-centric software engineering workflows.
By enabling rapid, iterative UML enrichment, LLM integration supports Agile and modern software engineering paradigms.

\section{Conclusion}
\label{sec:conclusion}

This study demonstrates the potential of large language models (LLMs) to automate the enrichment of UML class diagrams by generating method definitions directly from structured natural language use cases. Evaluating nine state-of-the-art LLMs across multiple dimensions—including method quantity, signature richness, annotation completeness, structural fidelity, syntactic correctness, and naming consensus—we provide a comprehensive performance assessment in the context of behavioral UML modeling.

Our results show that LLMs can produce methodically structured, class-specific methods that are syntactically correct and adhere to UML conventions. Models such as Claude~3.7 and ChatGPT-o3 excelled in method coverage and annotation completeness, while Mistral demonstrated richer parameterization. These findings reveal important trade-offs between comprehensive behavioral coverage and signature detail, emphasizing the need for careful model selection aligned with project goals.

Despite promising capabilities, inconsistencies in annotation completeness and signature richness underscore areas requiring further refinement. Improvements in prompt engineering, model fine-tuning, and ensemble strategies hold promise for enhancing output consistency and applicability.

The fast generation times and iterative refinement capabilities of LLMs align well with Agile software development practices, supporting rapid design cycles and adaptive workflows. By integrating LLM-assisted UML enrichment into software engineering pipelines, practitioners can accelerate early-stage modeling while maintaining traceability and structural integrity.

While this study focused primarily on UML class diagrams within the waste management domain, the presented methodology and analysis pipeline are broadly applicable across diverse software engineering fields. This generalizability opens avenues for evaluating LLM-generated models in other domains, such as healthcare, finance, and manufacturing, thereby enhancing the automation and quality assurance of software design practices universally.

All generated UML diagrams and associated metric datasets are publicly available, providing a valuable resource for reproducibility and benchmarking. Future work should focus on semantic validation, domain generalization, and human-AI collaboration protocols to advance trustworthy and practical AI-assisted software design.

\newpage
\section*{Declarations}

\begin{itemize}
\item \textbf{Funding:} The authors declare that they have no conflict of interest.
\item \textbf{Conflict of interest/Competing interests:} The authors declare that they have no conflict of interest.
\item \textbf{Ethics approval and consent to participate:} Not applicable. This study did not involve human participants, human data, or human tissue.
\item \textbf{Consent for publication:} All authors have given their consent for publication of this manuscript.
\begin{itemize}
  \item \textbf{Data availability:}  
  All data generated or analyzed during this study, including processed datasets, metric CSV files, and parsed JSON representations, are publicly available at the project GitHub repository: \url{https://github.com/Djaber1979/-Behavioral-Augmentation-of-UML-Class-Diagrams}.

  \item \textbf{Materials availability:}  
  All materials utilized in this research—including UML diagrams (\texttt{.puml}), rendered images (\texttt{.png}), and structured use cases—are accessible in the same GitHub repository.

  \item \textbf{Code availability:}  
  All code scripts for prompt generation, parsing, and metric computation are also provided at \url{https://github.com/Djaber1979/-Behavioral-Augmentation-of-UML-Class-Diagrams}, supporting reproducibility and further experimentation.
\end{itemize}

\item \textbf{Author contribution:} 
\begin{itemize}
 \item Djaber Rouabhia conceived and designed the study, conducted the experiments and wrote the manuscript.
 \item Ismail Hadjadj analyzed the data and reviewed the manuscript.
\end{itemize}
\end{itemize}

\bigskip
\begin{flushleft}%
Editorial Policies for:

\bigskip\noindent
Springer journals and proceedings: \url{https://www.springer.com/gp/editorial-policies}

\bigskip\noindent
Nature Portfolio journals: \url{https://www.nature.com/nature-research/editorial-policies}

\bigskip\noindent
\textit{Scientific Reports}: \url{https://www.nature.com/srep/journal-policies/editorial-policies}

\bigskip\noindent
BMC journals: \url{https://www.biomedcentral.com/getpublished/editorial-policies}
\end{flushleft}
\bibliography{sn-bibliography}

\newpage
\appendix
\section{Prompt Design}
\label{appendix:prompt}

The following is the full instruction prompt provided to each LLM, prior to presenting the UML diagram and structured use cases. It was used consistently across all models and runs:

\begin{quote}\small
\textbf{Objective} \\[6pt]

Analyze the UML class diagram (that will be provided in the next prompt) and main scenarios to refine/add domain-layer methods. Decompose each scenario into atomic business logic actions (state rules, entity interactions), ensuring strict adherence to SRP, and excluding UI/infrastructure concerns.

\vspace{8pt}
\textbf{Guidelines}  
\begin{enumerate}[leftmargin=*, label=\arabic*.]
    \item \textbf{Method Design}  
    \begin{itemize}[leftmargin=*, noitemsep]
        \item Create only methods addressing uncovered actions  
        \item Assign methods exclusively to classes owning the corresponding responsibility (SRP enforcement)  
        \item Modify existing methods \textit{only} if responsibilities exactly match  
    \end{itemize}

    \item  \textbf{Action Mapping}  
    \begin{itemize}[leftmargin=*, noitemsep]
        \item 1 method = 1 atomic business action (split at “And” clauses)  
        \item Allow technical sub-actions \textit{only} when serving a single business purpose  
    \end{itemize}

    \item  \textbf{State Management}  
    \begin{itemize}[leftmargin=*, noitemsep]
        \item Implement explicit state transitions (e.g., \texttt{confirmOrder()})  
        \item Embed state rules directly in methods; prohibit generic setters  
    \end{itemize}

    \item \textbf{Naming \& Structure}  
    \begin{itemize}[leftmargin=*, noitemsep]
        \item Apply domain verbs reflecting class responsibilities (e.g., \texttt{calculateTotal()}, never presentation terms)  
        \item Respect UML naming conventions and visibility modifiers (`-`, `+`, `\#`, `~`)  
        \item Maintain original UML notation/style/structure  
    \end{itemize}
\end{enumerate}

\vspace{8pt}
\textbf{Constraints} 
\begin{itemize}[leftmargin=*, noitemsep]
    \item Await class diagram \& UC specifications before analysis; do not proceed with assumptions.  
    \item Process \textit{only} primary scenarios  
    \item Preserve original class diagram structure (relationships/classes/attributes/enums/enum elements)  
    \item Update PlantUML iteratively (only add/modify methods + \texttt{//Uc\_id} + \texttt{//action: description})  
    \item Never move responsibilities between classes  
    \item \textbf{Final output:} Full PlantUML code including (original structure + added/modified methods)  
\end{itemize}

\vspace{8pt}
\textbf{Validation}  
\begin{itemize}[leftmargin=*, noitemsep]
    \item Strict 1:1 action-method mapping  
    \item All classes/methods comply with SRP  
    \item Zero UI/infrastructure terminology  
    \item Explicit state transition enforcement  
    \item Signature/naming consistency  
\end{itemize}

\vspace{6pt}
(No explanations or commentary)
\end{quote}

\section{LLM Model Specifications}
\label{appendix:llm-summary}

Table~\ref{tab:llm-summary} summarizes key properties of the nine LLMs used in this study, including version, release date, maximum context length, architectural details, and parameter counts (where available). This contextual information supports interpretation of performance differences observed in the main evaluation.

\begin{table}[ht]
\centering
\caption{Summary of model versions, release dates, context lengths, architectures, and parameter counts. Parameter estimates for proprietary models remain undisclosed.}
\label{tab:llm-summary}
\small
\begin{tabular}{|p{2.3cm}|p{1.3cm}|p{1.4cm}|p{5.6cm}|p{2.7cm}|}
\hline
\textbf{Model (Version)} & \textbf{Release Date} & \textbf{Max Context} & \textbf{Architecture / Training Highlights} & \textbf{Parameters (Total / Active)} \\
\hline
Claude 3.7 & Feb 2025 \cite{claude} & 200K (500K enterprise) & Dense LLM with hybrid reasoning (instant vs extended). Includes CLI assistant; RLHF-tuned for alignment \cite{claude}. & Undisclosed \\
\hline
Qwen 3 & Apr 2025 \cite{qwen} & $\sim$38K tokens & Hybrid dense/MoE model with multilingual and agent support; MoE variant (235B total, 22B active) \cite{qwen}. & Dense: 0.6B–32B; MoE: 235B (22B active) \\
\hline
ChatGPT-o3 & Apr 2025 \cite{chatgptO3} & 200K & OpenAI’s reasoning model for ChatGPT. RLHF-trained, excels in complex queries \cite{chatgptO3}. & Undisclosed \\
\hline
Gemini 2.5 Pro & May 2025 \cite{gemini} & 1M & Multimodal dense decoder; chain-of-thought and RLHF-tuned. Strong coding/multimodal tasks \cite{gemini}. & Undisclosed \\
\hline
DeepSeek-R1 & Jan 2025 \cite{deepseek} & 163K & Sparse MoE (128 experts); 671B total, 37B active per token. Trained for reasoning \cite{deepseek}. & 671B / 37B \\
\hline
Le Chat (Mistral Medium 3) & May 2025 \cite{mistralLeChat} & 128K & Dense autoregressive LLM; multimodal. Fine-tuned for long context and enterprise usage \cite{mistralLeChat}. & Undisclosed \\
\hline
GPT-4o & May 2024 \cite{gpt4o} & 128K & Multimodal dense decoder (text/audio/image/video); RLHF-trained; matches GPT-4 Turbo on text \cite{gpt4o}. & Undisclosed \\
\hline
Llama 4 (Scout / Maverick) & Apr 2025 \cite{llama} & 1M–10M & MoE LLMs; Scout (109B total, 17B active), Maverick (400B total, 17B active). Multimodal with CoT/RLHF \cite{llama}. & Scout: 109B / 17B; Maverick: 400B / 17B \\
\hline
\end{tabular}
\end{table}

\section{Metric Definitions, Symbols, and Context}
\label{sec:appendix-metrics}

This appendix formalises each metric applied in the study. Each entry includes:
\begin{enumerate}
 \item a brief \emph{definition},
 \item the corresponding \emph{formula}, and
 \item the specific \emph{context} of its use.
\end{enumerate}

These metrics correspond directly to the quantitative dimensions analyzed in the Results section, providing a formal foundation for measuring UML behavioral enrichment quality.

\subsection{Method Generation and Structural Fidelity}

\paragraph{\textbf{Method Quantity (MQ)}}  
\emph{Definition:} Total count of methods generated per UML class diagram.  
\[
 \text{MQ} = \sum_{i=1}^{C} M_i
\]
where \(M_i\) is the number of methods in class \(i\), and \(C\) is the total number of classes in the diagram.

\emph{Context:} Measures the extent of behavioral enrichment in terms of method coverage.

\paragraph{\textbf{Signature Richness (SR)}}  
\emph{Definition:} Composite metric capturing detail in method signatures, including average parameter count, proportion of non-void return types, and lexical diversity of method names.  
\[
 \text{SR} = \bigg( \overline{P}, \quad \text{RTC}, \quad D \bigg)
\]
where \(\overline{P}\) is the mean parameter count per method, \(\text{RTC}\) is return-type completeness (proportion non-void returns), and \(D\) is normalized Levenshtein distance among unique method names (see below).

\emph{Context:} Reflects complexity and informativeness of generated methods.

\paragraph{\textbf{Name Redundancy}}  
\emph{Definition:} Degree of name duplication in a method list.  
\[
 \text{Redundancy} = \frac{M}{U}, \quad
 M = \text{total methods}, \quad U = \text{unique names}.
\]  
\emph{Context:} High redundancy implies limited lexical variation in method names.

\paragraph{\textbf{Return-Type Completeness}}  
\emph{Definition:} Proportion of methods with explicit non-\texttt{void} return types.  
\[
 \text{RTC} = \frac{R}{M}, \quad
 R = \#\text{methods with } r \neq \texttt{void}.
\]  
\emph{Context:} Reflects informativeness of generated method signatures.

\paragraph{\textbf{Normalized Levenshtein Distance}}  
\emph{Definition:} Average edit distance between method names, normalized to [0,1].  
\[
 D = \frac{1}{\binom{U}{2}} \sum_{i<j} \frac{\text{Lev}(s_i, s_j)}{\max(|s_i|, |s_j|)}.
\]  
\emph{Context:} Lower values indicate greater lexical consistency in naming.

\subsection{Annotation Completeness and Method Consensus}

\paragraph{\textbf{Annotation Coverage}}  
\emph{Definition:} Proportion of methods containing explicit use-case or action annotations.  
\[
 \text{uccov} = \frac{M_{\text{UC}}}{M}, \quad
 \text{Actcov} = \frac{M_{\text{act}}}{M}.
\]  
\emph{Context:} Indicates traceability of generated behavior to original use cases.

\paragraph{\textbf{Core-Method Consensus}}  
\emph{Definition:} Proportion of LLMs that independently generated a core method.  
\[
 \text{Cons}(m) = \frac{\#\text{LLMs generating } m}{M_{\text{LLMs}}} \times 100.
\]  
\emph{Context:} Used to identify high-agreement methods across models.

\paragraph{\textbf{Structural Placement Consistency (SPC)}}  
\emph{Definition:} Proportion of LLMs that assign a method to the same UML class.  
\[
 \text{SPC}(m) = \frac{\#\text{LLMs placing } m \text{ in dominant class}}{\#\text{LLMs generating } m} \times 100.
\]  
\emph{Context:} Measures architectural consensus in behavioral enrichment.

\paragraph{\textbf{Jaccard Similarity}}  
\emph{Definition:} Ratio of the intersection to the union of two sets.  
\[
 J(A,B) = \frac{|A \cap B|}{|A \cup B|}.
\]  
\emph{Context:} Used to compare model-wise overlap in method sets.

\paragraph{\textbf{Syntactic Correctness (SC)}}  
\emph{Definition:} Proportion of generated UML diagrams that compile successfully without syntax errors.  
\[
 \text{SC} = \frac{\#\text{valid diagrams}}{\#\text{total diagrams}}.
\]  
\emph{Context:} Measures validity and toolchain compatibility of generated PlantUML code.

\paragraph{\textbf{Top-Method Consensus (TMC)}}  
\emph{Definition:} Lexical overlap of the most frequent normalized method names across models.  
\[
 \text{TMC} = \frac{|\bigcap_{m=1}^M S_m|}{|\bigcup_{m=1}^M S_m|}
\]  
where \(S_m\) is the set of top frequent method names for model \(m\), normalized to a canonical form.

\emph{Context:} Assesses lexical convergence and naming consistency across LLM outputs.

\section{Inferential and Non-Parametric Statistics}
\label{sec:appendix-stats}

This appendix details the inferential and non-parametric statistical tests used in the analysis. These tests complement the metric analyses described above by assessing statistical significance and effect sizes of observed differences. Each entry includes:  
\begin{enumerate}
  \item a brief \emph{definition},
  \item the corresponding \emph{formula}, and
  \item the specific \emph{context} of its use in this study.
\end{enumerate}

\subsection{Kruskal–Wallis Test}

\begin{enumerate}
  \item \emph{Definition:}  
  The Kruskal–Wallis test is a rank-based non-parametric test used to determine whether there are statistically significant differences between the medians of three or more independent groups. It is an extension of the Mann–Whitney U test when comparing more than two groups.

  \item \emph{Formula:}  
  Let \(N\) be the total number of observations, \(k\) the number of groups, and \(R_i\) the sum of ranks for group \(i\) with size \(n_i\). The test statistic \(H\) is:
  \[
    H = \frac{12}{N(N+1)} \sum_{i=1}^k \frac{R_i^2}{n_i} - 3(N+1)
  \]
  Under the null hypothesis, \(H\) approximates a chi-squared distribution with \(k-1\) degrees of freedom.

  \item \emph{Context:}  
  Used to test for global differences across LLM models in metrics such as Method Quantity, Signature Richness, Top-Method Consensus, and Core-Method Consensus where normality assumptions do not hold.
\end{enumerate}

\subsection{Dunn’s Post-Hoc Test}

\begin{enumerate}
  \item \emph{Definition:}  
  Dunn’s test is a non-parametric pairwise multiple comparison procedure performed following a significant Kruskal–Wallis test. It identifies which specific groups differ.

  \item \emph{Formula:}  
  For groups \(i\) and \(j\), the test statistic \(Z_{ij}\) is:
  \[
    Z_{ij} = \frac{R_i/n_i - R_j/n_j}{\sqrt{\frac{N(N+1)}{12} \left(\frac{1}{n_i} + \frac{1}{n_j}\right)}}
  \]
  where \(R_i\), \(n_i\) are the sum of ranks and size of group \(i\), respectively.

  \item \emph{Context:}  
  Applied post hoc to identify pairwise differences between LLMs in metrics where Kruskal–Wallis tests indicated significance, with Holm correction for multiple testing.
\end{enumerate}

\subsection{Chi-Squared Test of Independence}

\begin{enumerate}
  \item \emph{Definition:}  
  The chi-squared test assesses whether two categorical variables are independent by comparing observed and expected frequencies.

  \item \emph{Formula:}  
  \[
    \chi^2 = \sum_{i} \sum_{j} \frac{(O_{ij} - E_{ij})^2}{E_{ij}}
  \]
  where \(O_{ij}\) and \(E_{ij}\) are observed and expected frequencies for cell \((i,j)\).

  \item \emph{Context:}  
  Used to examine associations between categorical variables such as annotation completeness and model identity.
\end{enumerate}

\subsection{Wilcoxon Signed-Rank Test}

\begin{enumerate}
  \item \emph{Definition:}  
  The Wilcoxon signed-rank test is a non-parametric test for comparing two related samples to assess whether their population mean ranks differ.

  \item \emph{Formula:}  
  For paired differences \(d_i\), order the absolute values and assign ranks. Let \(W\) be the sum of ranks for positive differences. The test statistic is:
  \[
    W = \min(W^+, W^-)
  \]
  where \(W^+\) and \(W^-\) are sums of ranks for positive and negative differences, respectively.

  \item \emph{Context:}  
  Employed to assess Structural Placement Consistency across core methods between pairs of LLM models.
\end{enumerate}

\subsection{Cliff’s Delta (\(\delta\))}

\begin{enumerate}
  \item \emph{Definition:}  
  Cliff’s delta measures the effect size by quantifying the degree of overlap between two ordinal distributions. It ranges from \(-1\) to \(+1\), where 0 indicates no difference.

  \item \emph{Formula:}  
  Given two groups \(X\) and \(Y\),  
  \[
    \delta = \frac{\text{number of }(x_i > y_j) - \text{number of }(x_i < y_j)}{n_X n_Y}
  \]
  where \(n_X\) and \(n_Y\) are sample sizes of groups \(X\) and \(Y\).

  \item \emph{Context:}  
  Reported alongside Kruskal–Wallis and Dunn’s tests to indicate the magnitude of differences between LLMs in metrics such as method counts and parameter richness.
\end{enumerate}

\subsection{Cramér’s V}

\begin{enumerate}
  \item \emph{Definition:}  
  Cramér’s V measures the strength of association between two nominal variables, scaled between 0 (no association) and 1 (perfect association).

  \item \emph{Formula:}  
  \[
    V = \sqrt{\frac{\chi^2}{N \times (k - 1)}}
  \]
  where \(\chi^2\) is the chi-squared statistic, \(N\) is total observations, and \(k\) is the smaller number of categories in the variables.

  \item \emph{Context:}  
  Used to quantify effect size in chi-squared tests, such as the association between model identity and annotation completeness.
\end{enumerate}

\newpage
\section{Placement consistency}
\label{appendix:spc-core-methods}
\begin{table}[ht]
\centering
\caption{Placement consistency of core methods across LLMs: method name, dominant class, number of LLMs agreeing on generating the method, number of LLMs agreeing on placement in the dominant class, and consistency rate (\%).}
\label{tab:spc-core-methods}
\small
\begin{tabular}{l l c c c}
\toprule
\textbf{Method} & \textbf{Dominant Class} & \textbf{\#LLMs Agreeing} & \textbf{\#LLMs Agreeing on Placement} & \textbf{Consistency (\%)} \\
\midrule
updatePassword             & UserCredentials     & 9  & 9  & 100.0 \\
cancelRequest             & ServiceRequest      & 9  & 9  & 100.0 \\
assignRole                & RoleManager         & 9  & 9  & 100.0 \\
resetPassword             & UserCredentials     & 9  & 9  & 100.0 \\
verifyEmail               & UserProfile         & 9  & 8  & 88.9  \\
lockaccount               & Useraccount         & 8  & 8  & 100.0 \\
deactivateaccount         & Useraccount         & 8  & 8  & 100.0 \\
sendVerificationEmail     & NotificationService & 8  & 8  & 100.0 \\
generatePasswordResetToken & UserCredentials     & 8  & 8  & 100.0 \\
validateCredentials       & UserCredentials     & 8  & 8  & 100.0 \\
updateProfile             & UserProfile         & 8  & 8  & 100.0 \\
cancel                    & ServiceRequest      & 8  & 8  & 100.0 \\
deactivate                & Useraccount         & 7  & 7  & 100.0 \\
activateaccount           & Useraccount         & 7  & 7  & 100.0 \\
validateRoleTransition    & RoleManager         & 7  & 7  & 100.0 \\
sendNotification          & NotificationService & 6  & 6  & 100.0 \\
schedule                  & ServiceRequest      & 6  & 6  & 100.0 \\
updateListing             & Product             & 6  & 6  & 100.0 \\
createListing             & Product             & 6  & 6  & 100.0 \\
addStage                  & WasteJourney        & 6  & 6  & 100.0 \\
listProduct               & Product             & 4  & 4  & 100.0 \\
applyFilters              & SearchFilter        & 4  & 4  & 100.0 \\
sendPasswordResetConfirmation & NotificationService & 3  & 3  & 100.0 \\
updateStatus              & ServiceRequest      & 7  & 6  & 85.7  \\
scheduleTransport         & TransportRequest    & 7  & 6  & 85.7  \\
updateDetails             & CollectionPoint     & 8  & 6  & 75.0  \\
login                     & Useraccount         & 6  & 5  & 83.3  \\
updateProfileDetails      & UserProfile         & 6  & 5  & 83.3  \\
validateEmailFormat       & UserProfile         & 6  & 5  & 83.3  \\
scheduleCollection        & CollectionRequest   & 6  & 5  & 83.3  \\
updateLocation            & WasteJourney        & 7  & 5  & 71.4  \\
recordLogin               & Useraccount         & 5  & 4  & 80.0  \\
acceptRequest             & CollectionRequest   & 8  & 4  & 50.0  \\
validate                  & Address             & 5  & 3  & 60.0  \\
createaccount             & Useraccount         & 7  & 7  & 100.0 \\
publishListing            & Product             & 7  & 7  & 100.0 \\
\bottomrule
\end{tabular}
\end{table}

\end{document}